\documentclass[a4paper,11pt]{article}

\usepackage[T1]{fontenc} 
\usepackage[utf8]{inputenc}

\usepackage{jheppub,bm,mathtools} 
\usepackage[usenames,dvipsnames,svgnames,table,x11names]{xcolor}

\renewcommand*{\backref}[1]{}
\renewcommand*{\backrefalt}[4]{%
  \ifcase #1 %
    \relax 
  \or
    {\scriptsize (page~#2).}%
  \else
    {\scriptsize (pages~#2).}%
  \fi%
}

\definecolor{light_blue}{rgb}{0.15, 0.35, 0.95}
\definecolor{kit_green}{rgb}{
    0, 
    0.58823, 
    0.50980  
}

\usepackage{tikz}
\usepackage{pgfplots}
\pgfplotsset{compat=1.14}
\usepackage{tikz-3dplot}
\usetikzlibrary{intersections, calc,positioning,decorations.pathreplacing,decorations.pathmorphing,arrows,arrows.meta,patterns,pgfplots.fillbetween,math,matrix}

\DeclareMathOperator{\disc}{Disc}
\DeclareMathOperator{\im}{Im}
\DeclareMathOperator*{\res}{Res}

\newcommand{\A}{\mathcal{A}}
\newcommand{\Mpl}{M_\text{Pl}}

\DeclareMathOperator{\SO}{SO}
\DeclareMathOperator{\U}{U}

\renewcommand{\d}[2][]{\mathrm{d}^{#1}{#2}}

\usepackage{letltxmacro}
\makeatletter
\let\oldr@@t\r@@t
\def\r@@t#1#2{%
\setbox0=\hbox{$\oldr@@t#1{#2\,}$}\dimen0=\ht0
\advance\dimen0-0.2\ht0
\setbox2=\hbox{\vrule height\ht0 depth -\dimen0}%
{\box0\lower0.4pt\box2}}
\LetLtxMacro{\oldsqrt}{\sqrt}
\renewcommand*{\sqrt}[2][\ ]{\oldsqrt[#1]{#2}}
\makeatother

\begin{document}

\title{
    On (Scalar QED) Gravitational Positivity Bounds
}

\author{
    \textbf{Yuta Hamada}$^{\diamond\star}\,$,\,
    \textbf{Rinto Kuramochi}$^\star\,$,\,
    \textbf{Gregory J.\ Loges}$^\diamond\,$,\,
     \textbf{Sota Nakajima}$^\diamond\,$
    \vspace{5mm}
    \\
    $^\diamond$\normalsize{\it Theory Center, IPNS, High Energy Accelerator Research Organization (KEK),}\\
    \normalsize{\it\quad 1-1 Oho, Tsukuba, Ibaraki 305-0801, Japan} \\
    $^\star$\normalsize{\it Graduate University for Advanced Studies (Sokendai), 1-1 Oho, Tsukuba, }
    \\
    \normalsize{\it\quad Ibaraki 305-0801, Japan}\vspace{1mm} 
}

\preprint{KEK-TH-2492}

\emailAdd{yhamada@post.kek.jp}
\emailAdd{rinto@post.kek.jp}
\emailAdd{gloges@post.kek.jp}
\emailAdd{snakajim@post.kek.jp}

\abstract{\noindent\normalsize
    We study positivity bounds in the presence of gravity.
    We first review the gravitational positivity bound at the tree-level, where it is known that a certain amount of negativity is allowed for the coefficients of higher-derivative operators.
    The size of these potentially negative contributions is estimated for several tree-level, Reggeized gravitational amplitudes which are unitary at high energies and feature the $t$-channel pole characteristic of graviton exchange.
    We also argue for the form of the one-loop Regge amplitude assuming that the branch cut structure associated with the exchange of the graviton and higher-spin particles is reflected.
    We demonstrate how the one-loop Regge amplitude appears by summing over Feynman diagrams.
    For our one-loop amplitude proposal, the positivity bounds generically receive a finite contribution from the Regge tower and do not lead to a parametrically small bound on the cut-off scale of the low-energy EFT, consistent with recent studies based on sum rules of the amplitude.
}

\maketitle


\section{Introduction}
\label{sec:intro}

The space of low-energy EFTs is highly constrained by unitarity, locality, and causality~\cite{Adams:2006sv}. Imposing these general requirements allows one to bound, for example, the Wilson coefficients of higher-derivative operators and rule out a whole swathe of theories as being inconsistent.
For instance, the optical theorem on the scattering amplitude in the forward limit leads to a variety of positivity bounds on the Wilson coefficients.
Recent developments reveal that the infinite tower of higher-dimension operators must lie inside the EFThedron~\cite{Arkani-Hamed:2020blm,Chiang:2021ziz,Chiang:2022ltp}.

The Swampland program \cite{Vafa:2005ui} (see~\cite{Palti:2019,vanBeest:2021lhn,Grana:2021zvf,Agmon:2022thq} for recent reviews) takes this one step further by aiming to understand the nontrivial imprints that quantum gravity has on low-energy physics.
There are many conjectures such as the Weak Gravity Conjecture~\cite{Arkani-Hamed:2006emk}, Distance Conjecture~\cite{Ooguri:2006in}, and so forth.
It is desirable to provide explanations for these conjectures.

It is natural to utilize the technique of the positivity bound to explain the Swampland conjectures.
This idea is particularly useful to demonstrate the Weak Gravity Conjecture, which states the existence of the state where the charge is larger than the mass.\footnote{In addition to the Weak Gravity Conjecture, the S-matrix bootstrap has been used to obtain a lower bound on the coefficient of the higher-derivative operators~\cite{Guerrieri:2021ivu,Bern:2021ppb,Bern:2022yes}.} The certain positivity condition on the four derivative operators means that the Weak Gravity Conjecture is realized by the nearly extremal black holes~\cite{Kats:2006xp,Cheung:2018cwt,Hamada:2018dde,Bellazzini:2019xts,Loges:2019jzs,Goon:2019faz,Jones:2019nev,Loges:2020trf,Arkani-Hamed:2021ajd,Cao:2022iqh}.
However, in contrast to the non-gravitational case, it is difficult to obtain positivity bounds for gravitational amplitudes.
The main obstruction comes from the existence of the massless graviton, which leads to the $t$-channel pole at $t=0$.
Because of this, one cannot directly take the forward limit, $t\to0$, where the optical theorem is applicable.
Ref.~\cite{Tokuda:2020mlf} bypasses this difficulty\footnote{Another direction to study the gravitational positivity bound is to work with a fixed, finite impact parameter~\cite{Caron-Huot:2021rmr}, which works for $d\geq5$.} by assuming a Regge form for the amplitude at high energy, and shows that the gravitational amplitude satisfies an approximate positivity bound.\footnote{See also Ref.~\cite{Hamada:2018dde} for the schematic idea of the approximate positivity bound.}
The Regge amplitude means that, for $s\to\infty$ with fixed $t<0$, the amplitude $\A$ satisfies $\lim_{s\to\infty}\A/s^2=0$.\footnote{The argument based on the causality~\cite{Arkani-Hamed:2020blm} indicates $\lim_{s\to\infty}\A/s^2=\text{(finite)}$, though it is hard to say that the finite value is zero. 
The chaos bound~\cite{Maldacena:2015waa,Chandorkar:2021viw} leads to the same conclusion. 
It is also argued that the Regge boundedness for $d\geq5$ follows from the reasonable assumptions in Ref.~\cite{Haring:2022cyf}.}
This assumption is mainly motivated by the behavior of the Virasoro-Shapiro or heterotic string amplitude in tree-level string theory.
Moreover, other amplitudes satisfying Regge boundedness and the IR consistency conditions such as unitarity have been found~\cite{Cheung:2022mkw,Geiser:2022exp}.

Recently, the gravitational positivity bound at the one-loop level has been used to
speculate new Swampland conjectures. 
It was argued in Ref.~\cite{Alberte:2020bdz} that for scalar QED coupled to Einstein gravity, the gravitational positivity bounds provide a nontrivial relationship between the EFT cut-off $\Lambda$ and the scalar's mass $m$ and charge $e$,
\begin{equation}\label{eq:lowCutOff}
    \Lambda \lesssim \sqrt{em\Mpl} \,,
\end{equation}
assuming that the massless graviton $t$-channel pole may be discarded.\footnote{The bound is still valid after taking into account the effect of the $t$-channel pole~\cite{Tokuda:2020mlf,Aoki:2021ckh} as long as we assume the Reggeized amplitude motivated by the tree-level string theory. See Sec.~\ref{Sec:Cancellation} and Sec.~\ref{Sec:small_cutoff} for the detail.} Here $\Mpl$ is the reduced Planck mass.
This scale is much smaller than expected and is even parametrically smaller than that provided by the magnetic Weak Gravity Conjecture~\cite{Arkani-Hamed:2006emk},
\begin{equation}\label{eq:magWGC}
    \Lambda \lesssim e\Mpl \,,
\end{equation}
when the scalar is light, $m\ll e\Mpl$.
A similar technique is used to obtain the cut-off scale of the Standard Model~\cite{Aoki:2021ckh} and dark photon model~\cite{Noumi:2022zht} as well as constraining the shape of the scalar potential~\cite{Noumi:2021uuv}.

As stressed in Ref.~\cite{Alberte:2020bdz}, the bound~\eqref{eq:lowCutOff} does not hold if the graviton $t$-channel pole can not be discarded. 
This is roughly equivalent to the question of whether we can neglect the contributions from the Regge states or not.
This is because, in addition to canceling the $t$-channel pole, the Regge states can contribute a subleading $\mathcal{O}(t^0s^2)$ term to the amplitude which has a nontrivial effect on the positivity bounds in the low-energy EFT~\cite{Tokuda:2020mlf}. 
Ref.~\cite{Alberte:2021dnj} (and very recently \cite{deRham:2022gfe} in the case of the graviton-graviton scattering) argues that the contributions from the Regge states can not be neglected for the graviton-photon scattering.
Given an IR amplitude, sum rules indicate that the amplitude from Regge states alters the argument which leads to Eq.~\eqref{eq:lowCutOff}.\footnote{See also Ref.~\cite{Herrero-Valea:2022lfd} for the study of the Regge amplitude including the effect of the graviton loop. Ref.~\cite{Tokuda:2020mlf} puts the bound on the form of the Regge amplitude by using the finite energy sum rule and IR part of null constraints.}
Moreover, in this paper, we provide a counterexample of the supersymmetric version of Eq.~\eqref{eq:lowCutOff} assuming the IR consistency and the Regge boundedness of the amplitude.
This is an independent argument that graviton $t$
-channel pole can not be neglected in the case of photon-photon amplitude.

If this is the case, then it is likely that the high energy Regge amplitude contains the light mass parameter $m^2$ in the denominator of the prefactor.\footnote{Another option is that the Regge tower is light, but we will argue that this option is unlikely (at least for some cases) given our counterexample.} 
However, since it is expected that the Reggeization occurs due to the effect of the heavy higher spin states, it is not obvious why the amplitude with the light mass parameter appears.
In this article, we demonstrate that this can be understood by accounting for 1-loop effects on the amplitude from $t$-channel exchange of Regge states. 

We also study the properties of `deformed' tree-level gravitational amplitudes.
A subleading $\mathcal{O}(t^0s^2)$ term to the amplitude is computed, which has an impact on the positivity bounds.

\medskip

The remainder of this article is organized as follows. In Sec.~\ref{sec:treelevel} we review the positivity bounds for $2$-to-$2$ photon scattering in non-gravitational theories and how, when minimally coupled to Einstein gravity, the resulting $t$-channel pole is canceled by Reggeization. In Sec.~\ref{sec:deformed} we analyze the properties of `deformed' tree-level gravitational amplitudes which have been recently proposed. These amplitudes satisfy the IR consistency conditions, and we study the implications of these amplitudes on the positivity bound. In Sec.~\ref{sec:oneloop} we consider the subleading, 1-loop contributions to scalar QED and show that the Regge states lead to a nonzero correction in the forward limit, $t\to0$, assuming that the branch cut structure associated with the exchange of the graviton and higher spin particles is reflected in the Regge amplitude.
We conclude with a discussion in Sec.~\ref{sec:discussion}.


\section{Positivity bounds at tree level}
\label{sec:treelevel}

In this section we review how gravitational positivity bounds may be extracted in the forward limit $t\to0^-$ despite the presence of the $t$-channel pole coming from graviton exchange.
Reggeization of the gravitational amplitudes provides a mechanism for canceling the $s^2/t$ term that appears in Einstein gravity. After carefully subtracting such divergent terms, dispersive sum rules allow one to derive positivity bounds on the remaining finite terms which are calculable in a low-energy EFT.

We will focus on the fixed-$t$ scattering of two photons where all incoming and outgoing particles have $+$ helicity. The amplitudes can be written in terms of the usual Mandelstam variables $s,t,u$ which satisfy $s+t+u=0$. Later in Sec.~\ref{Sec:loop}, for concreteness, we will consider scalar QED coupled to Einstein gravity in 4d, for which the quadratic action is
\begin{equation}\label{eq:scalarQEDaction}
    S = \int\d[4]{x}\,\sqrt{-g}\left(\frac{\Mpl^2}{2}\,R - \frac{1}{4}F_{\mu\nu}F^{\mu\nu} - D_\mu\phi^\dagger D^\mu\phi - m^2\phi^\dagger\phi\right) \,,
\end{equation}
where the mostly-plus notation is adopted. See Sec.~\ref{sec:notation} for more details on our conventions and notation.

\subsection{Review: (non-)gravitational positivity bounds}

Let us begin by recalling the usual argument which allows one to derive positivity bounds in EFTs, taking care of the difference between the non-gravitational and gravitational amplitudes. Along the way we note the points where the argument cannot immediately be adopted for theories coupled to Einstein gravity.

In discussing dispersion relations we make the usual Regge-boundedness assumption on the high-energy behavior of the amplitude:
\begin{equation}\label{eq:s2bounded}
    \lim_{\substack{|s|\to\infty\\ t<0\text{ fixed}}}\frac{\A(s,t)}{s^2} = 0 \,.
\end{equation}
For the non-gravitational case, this is guaranteed by the Froissart theorem~\cite{Froissart:1961ux,Martin:1962rt} in a gapped theory.
For the gravitational case, this is of course violated by tree-level graviton exchanged, but satisfied when the graviton is Reggeized, as we will consider below. We also assume that $\A$ is crossing symmetric and holomorphic in $s$, except for poles and cuts along the real axis.

\begin{figure}[t]
    \centering
    \definecolor{myRed}{rgb}{0.7,0,0}
    \definecolor{myBlue}{rgb}{0,0,0.5}
    \definecolor{myGreen}{rgb}{0,0.5,0}
    \newcommand{\w}{0.07}
    \begin{tikzpicture}
        \draw[thick, <->] (-6,0) -- (6,0);
        \draw[thick, <->] (0,-3) -- (0,3);
        \node[draw=black] at (5.5,2.5) {$s$};

        \draw[ultra thick, decorate, decoration={zigzag}, myBlue] (3,0) -- (5.8,0);
        \draw[ultra thick, decorate, decoration={zigzag}, myBlue] (-3,0) -- (-5.8,0);
        \fill[myBlue] (3,0) circle (0.06);
        \fill[myBlue] (-3,0) circle (0.06);

        \foreach\x in {-2,-1.3,0,1.3,2}{
            \draw[very thick, myBlue] (\x-\w,-\w) -- (\x+\w,\w);
            \draw[very thick, myBlue] (\x-\w,\w) -- (\x+\w,-\w);
        };

        \foreach\th in {0,45,90,135,180,225,270,315}{
            \draw[myGreen!35, ->] (1,1.2) -- ({1+0.6*cos(\th)},{1.2+0.6*sin(\th)});
        };
        \fill[white] (1,1.2) circle (0.3);
        \draw[thick, myGreen, ->] (1.2,1.2) arc (0:360:0.2);
        \node[above] at (1.4,1.05) {$\gamma$};
        
        \draw[thick, dashed, myRed] (5.5,0.3) arc (0:180:5.5 and 2.25);
        \draw[thick, dashed, myRed] (5.5,-0.3) arc (360:180:5.5 and 2.25);
        \node[above] at (-1.5,-2.5) {$\mathcal{C}_\infty$};

        \draw[thick, myRed, ->] (0.3,0) arc (0:360:0.3);
        \node[above] at (0.4,0.05) {$\epsilon$};

        \draw[thick, myRed] (5.5,0.3) -- (3,0.3) arc (90:180:0.25) -- (2.2,0.05) arc (0:180:0.2 and 0.15) -- (1.5,0.05) arc (0:180:0.2 and 0.15) -- (1.1,-0.05) arc (180:360:0.2 and 0.15) -- (1.8,-0.05) arc (180:360:0.2 and 0.15) -- (2.75,-0.05) arc (180:270:0.25) -- (5.5,-0.3);
        \draw[->, myRed] (4,0.3) -- (4.05,0.3); 
        \draw[->, myRed] (4,-0.3) -- (3.95,-0.3);
        \node[above] at (3.5,0.3) {$\mathcal{C}_+$};
        \node[above] at (-3.5,0.3) {$\mathcal{C}_-$};
        \draw[thick, myRed] (-5.5,0.3) -- (-3,0.3) arc (90:0:0.25) -- (-2.2,0.05) arc (180:0:0.2 and 0.15) -- (-1.5,0.05) arc (180:0:0.2 and 0.15) -- (-1.1,-0.05) arc (360:180:0.2 and 0.15) -- (-1.8,-0.05) arc (360:180:0.2 and 0.15) -- (-2.75,-0.05) arc (360:270:0.25) -- (-5.5,-0.3);
        \draw[->, myRed] (-4,0.3) -- (-3.95,0.3); 
        \draw[->, myRed] (-4,-0.3) -- (-4.05,-0.3); 
    \end{tikzpicture}
    \caption{Analytic structure of $\A$ and the contour deformation of $\gamma$ into $\mathcal{C}_\infty + \mathcal{C}_+ + \mathcal{C}_- - \epsilon$.}
    \label{fig:contour_deformation}
\end{figure}
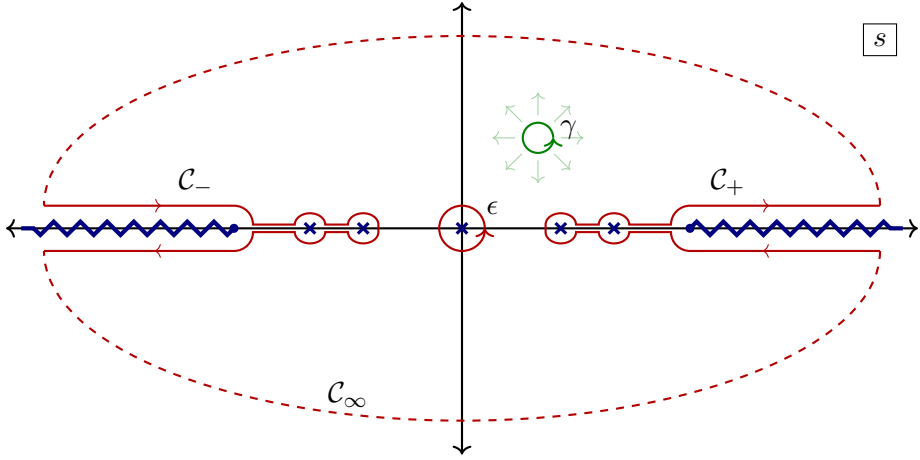

Start with the identity
\begin{equation}\label{eq:contour}
    \A(s,t) = s^2\oint_\gamma\frac{\d{s^\prime}}{2\pi i}\,\frac{\A(s^\prime,t)}{(s^\prime-s)s^{\prime2}} \,,
\end{equation}
where $\gamma$ is any contour encircling $s$ within which $\A$ is holomorphic. This integration contour can be deformed into a circle $\mathcal{C}_\infty$ at $|s^\prime|\to\infty$, contours $\mathcal{C}_\pm$ around any poles and branch cuts on the positive and negative real axes, and a contour $\epsilon$ which encircles the pole at $s=0$ (see Fig.~\ref{fig:contour_deformation}):
\begin{equation}\label{eq:contourdecomp}
    \A(s,t) = s^2\left( \int_{\mathcal{C}_\infty} + \int_{\mathcal{C}_+} + \int_{\mathcal{C}_-} -\oint_\epsilon \right)\frac{\d{s^\prime}}{2\pi i}\,\frac{\A(s^\prime,t)}{(s^\prime-s)s^{\prime2}} \,.
\end{equation}
The circle at infinity has vanishing contribution thanks to Eq.~\eqref{eq:s2bounded}. In the presence of gravity, $\A(s,t)$ may be expanded as
\begin{align}
    \A(s,t) = \frac{a_{-1}(t)}{s} + a_0(t) + a_1(t)s + \cdots
\end{align}
around the origin: note that the first term is absent without gravity. By substituting this into Eq.~\eqref{eq:contourdecomp}, the pole at the origin gives rise to
\begin{align}
    -s^2\oint_\epsilon\frac{\d{s^\prime}}{2\pi i}\,\frac{\A(s^\prime,t)}{(s^\prime-s)s^{\prime2}} 
    =\frac{a_{-1}(t)}{s}+a_0(t)+a_1(t)s \,.
\end{align}
This means that the scattering amplitude $\A$ can be written as\footnote{See Ref.~\cite{Sinha:2020win} for manifestly crossing symmetric dispersion relations.}
\begin{equation}\label{eq:dispersion0}
    \A(s,t,u) = \frac{a_{-1}}{s} + a_0 + a_1s + \frac{s^2}{\pi}\int_0^\infty\d{s^\prime}\,\frac{\disc_s\A(s^\prime,t,u^\prime)}{s^{\prime2}(s^\prime-s)} + \frac{u^2}{\pi}\int_0^\infty\d{u^\prime}\,\frac{\disc_u\A(s^\prime,t,u^\prime)}{u^{\prime2}(u^\prime-u)} \,,
\end{equation}
where the discontinuities $\disc_s$ and $\disc_u$ on the real axis are defined by\footnote{The discontinuity is related to the imaginary part by $\disc_s\A(s,t,u)=2i\im\A(s+i\epsilon,t,u)$.}
\begin{equation}
\begin{aligned}
    2i\disc_s\A(s,t,u) &:= \A(s+i\epsilon,t,u) - \A(s-i\epsilon,t,u) \,,\\
    2i\disc_u\A(s,t,u) &:= \A(s,t,u+i\epsilon) - \A(s,t,u-i\epsilon) \,.
\end{aligned}
\end{equation}
By comparing $s^2$ terms around $s=0$ in~\eqref{eq:dispersion0}, we obtain
\begin{equation}\label{eq:dispersion1}
    \partial_s^2\tilde{\A}(0,t,-t)
    =
    \frac{1}{\pi}\int_0^\infty\d{s^\prime}\,\frac{\disc_s\A(s^\prime,t,u^\prime)}{s^{\prime3}}
    +
    \frac{1}{\pi}\int_0^\infty\d{u^\prime}\,\frac{\disc_u\A(s^\prime,t,u^\prime)}{(u^\prime+t)^3}
\end{equation}
where we have defined
\begin{equation}
    \tilde{\A}:=\A-\frac{a_{-1}}{s}.
\end{equation}
This is an amplitude where the graviton $s$-channel pole is subtracted.

In the forward limit, the right-hand side of Eq.~\eqref{eq:dispersion1} must be non-negative by unitarity. This immediately gives the following positivity bounds,
\begin{equation}
    \partial_s^2\A(0,0,0) \geq 0 \,,
\end{equation}
which directly constrain the space of allowed EFTs. We can obtain stronger bounds by extracting the portion of the integrals along the cuts which are calculable at low energies. In order to do this, introduce $\Lambda$ as the EFT cut-off and write
\begin{equation}
\begin{aligned}
    \disc_s\A(s^\prime,t,u^\prime) &= \disc_s\A_\text{IR}(s^\prime,t,u^\prime)\,\theta(\Lambda^2 - s) + \disc_s\A_\text{UV}(s^\prime,t,u^\prime)\,\theta(s-\Lambda^2) \,,\\
    \disc_u\A(s^\prime,t,u^\prime) &= \disc_u\A_\text{IR}(s^\prime,t,u^\prime)\,\theta(\Lambda^2 - s) + \disc_u\A_\text{UV}(s^\prime,t,u^\prime)\,\theta(s-\Lambda^2) \,.
\end{aligned}
\end{equation}
Eq.~\eqref{eq:dispersion1} now becomes
\begin{equation}\label{eq:dispersion2}
\begin{aligned}
    &\hspace{-20pt} \partial_s^2\tilde{\A}(0,t,-t) - \frac{2}{\pi}\int_0^{\Lambda^2}\d{s^\prime}\,\frac{\disc_s\A_\text{IR}(s^\prime,t,u^\prime)}{s^{\prime3}} - \frac{2}{\pi}\int_0^{\Lambda^2}\d{u^\prime}\,\frac{\disc_u\A_\text{IR}(s^\prime,t,u^\prime)}{(u^\prime+t)^3}\\
    &= \frac{2}{\pi}\int_{\Lambda^2}^\infty\d{s^\prime}\,\frac{\disc_s\A_\text{UV}(s^\prime,t,u^\prime)}{s^{\prime3}} + \frac{2}{\pi}\int_{\Lambda^2}^\infty\d{u^\prime}\,\frac{\disc_u\A_\text{UV}(s^\prime,t,u^\prime)}{(u^\prime+t)^3} \,,
\end{aligned}
\end{equation}
where again the right-hand side is non-negative by unitarity in the forward limit. Without gravity there is no obstacle to taking $t\to0$, in which case the improved positivity bound reads~\cite{Bellazzini:2016xrt,deRham:2017imi,deRham:2017xox}
\begin{equation}\label{eq:improved}
    \partial_s^2\tilde{\A}(0,0,0) - \frac{2}{\pi}\int_0^{\Lambda^2}\d{s^\prime}\,\frac{\disc_s\A_\text{IR}(s^\prime,0,u^\prime)}{s^{\prime3}} - \frac{2}{\pi}\int_0^{\Lambda^2}\d{u^\prime}\,\frac{\disc_u\A_\text{IR}(s^\prime,0,u^\prime)}{u^{\prime3}} \geq 0 \,.
\end{equation}
By design everything here depends only on low-energy data and constrains the space of allowed EFTs, e.g.\ through constraining the Wilson coefficients of higher-derivative operators. One possibility is that the above equation provides an upper bound on $\Lambda$, the scale where new physics must appear to maintain unitarity.

\subsection{Cancellation of the \texorpdfstring{$\mathbf{t}$}{t}-channel pole in gravitational amplitudes}\label{Sec:Cancellation}

For gravitational systems the argument which leads to Eq.~\eqref{eq:improved} needs re-evaluating; more care is needed when graviton exchange is included because of the contribution
\begin{equation}
    \A_\text{grav}^\text{tree} \sim -\frac{s^2}{\Mpl^2t}
\end{equation}
which renders $\partial_s^2\tilde{\A}(0,0,0)$ ill-defined. Of course this means that the integrals on the right-hand side of~\eqref{eq:dispersion1} must diverge for $t\to0^-$ as well. 
In the literatures, there are two ways to deal with the graviton $t$-pole: 
\begin{enumerate}
    \item Subtracting the graviton $t$-pole assuming that the amplitude is Reggeized at the high energy~\cite{Tokuda:2020mlf}.
    \item Studying the gravitational positivity bound with a finite impact parameter~\cite{Caron-Huot:2021rmr,Caron-Huot:2022ugt,Caron-Huot:2022jli}.
\end{enumerate}
We focus on the first approach in this paper,\footnote{As for the second approach, the bound on the Wilson coefficients is obtained up to an infrared logarithm~\cite{Caron-Huot:2022ugt}. The implications on the black hole weak gravity conjecture are investigated in Ref.~\cite{Henriksson:2022oeu}.} 
 which was fleshed out in Ref.~\cite{Tokuda:2020mlf}; we assume that at energies well above the scale $M_\ast$ the amplitude is Reggeized,
\begin{equation}\label{eq:Regge_limit}
    \lim_{\substack{|s|\gg M_\ast^2\\ t<0\text{ fixed}}}\disc_s\A(s,t) 
    =f(t)\,\frac{s^{2+\alpha^\prime t}}{\Mpl^2M_\ast^{2\alpha^\prime t}} 
    \approx f_\text{tree}(t)\,\frac{s^{2+\alpha^\prime t}}{\Mpl^2M_\ast^{2\alpha^\prime t}} 
    =: \disc_s\A_\text{Regge}^\text{tree}(s,t) \,,
\end{equation}
where $\alpha^\prime$ provides the slope of the Regge trajectory, $f(t)$ is a function of $t$ whcih has mass dimension $-2$.
Here we have made the tree-level approximation of a UV theory, $f\approx f_\text{tree}$.
This is what happens in string theory, for example, with $M_\ast^2\approx M_\text{s}^2=\alpha^{\prime-1}$. Substituting this limiting expression into the dispersive integral gives
\begin{equation}\label{eq:tree_Regge}
    \int_{M_\ast^2}^\infty\d{s^\prime}\,\frac{\disc_s\A(s^\prime,t)}{s^{\prime3}} \approx \int_{M_\ast^2}^\infty\d{s^\prime}\,\frac{\disc_u\A(u^\prime,t)}{(u^\prime+t)^3}
    \approx
    -\frac{f_\text{tree}(t)}{\Mpl^2\alpha^\prime t} 
    =-\frac{f_\text{tree}(0)}{\Mpl^2\alpha^\prime t}
    -\frac{f_\text{tree}^\prime(0)}{\Mpl^2\alpha^\prime}
    +\cdots\,.
\end{equation}
As long as $f_\text{tree}(0)\sim\alpha^\prime$ is chosen appropriately, this divergent UV contribution to the dispersion relation cancels the $t$-channel pole in the IR. After the cancellation, the Regge contribution to $\disc\A$ need not be positive. For example, since at tree-level in string theory the function $f_\text{tree}(t)$ only depends on the dimensionful parameter $\alpha^\prime$, a certain amount of negativity is allowed:
\begin{align}
    \int_{M_\ast^2}^\infty\d{s^\prime}\,\frac{\disc_s\A(s^\prime,t)}{s^{\prime3}} \approx -\frac{1}{\Mpl^2t} - \mathcal{O}(1)\frac{\alpha^\prime}{\Mpl^2} + \mathcal{O}(t) \,.
\end{align}
The size of this negative $\mathcal{O}(t^0)$ term is crucial in discussing the possibility that (nearly) extremal black holes satisfy the Weak Gravity Conjecture~\cite{Arkani-Hamed:2006emk,Hamada:2018dde,Loges:2019jzs,Loges:2020trf,Arkani-Hamed:2021ajd}.

By substituting the tree-level Regge amplitude into the dispersion relation, we obtain
\begin{equation}\label{eq:dispersion3}
\begin{aligned}
    &\!\!\!\partial_s^2\Delta\A(0,t,-t) - \frac{2}{\pi}\int_0^{\Lambda^2}\d{s^\prime}\,\frac{\disc_s\A_\text{IR}(s^\prime,t,u^\prime)}{s^{\prime3}} - \frac{2}{\pi}\int_0^{\Lambda^2}\d{u^\prime}\,\frac{\disc_u\A_\text{IR}(s^\prime,t,u^\prime)}{(u^\prime+t)^3}\\
    &=
    \frac{2}{\pi}\left(\int_{\Lambda^2}^{M_*^2}\d{s^\prime}\,\frac{\disc_s\A_\text{UV}(s^\prime,t,u^\prime)}{s^{\prime3}} +\int_{\Lambda^2}^{M_*^2}\d{u^\prime}\,\frac{\disc_u\A_\text{UV}(s^\prime,t,u^\prime)}{(u^\prime+t)^3}\right)\\
    &+\frac{2}{\pi}\left(\int^{\infty}_{M_*^2}\d{s^\prime}\,\frac{\disc_s\Delta\A_\text{UV}(s^\prime,t,u^\prime)}{s^{\prime3}} +\int^{\infty}_{M_*^2}\d{u^\prime}\,\frac{\disc_u\Delta\A_\text{UV}(s^\prime,t,u^\prime)}{(u^\prime+t)^3}\right)
    -\frac{2f_\text{tree}^\prime(t)}{\Mpl^2\alpha^\prime},
\end{aligned}
\end{equation}
where 
\begin{equation}
    \Delta\A:=\tilde{\A}-\A_\text{grav}^\text{tree} \,, \qquad \Delta\A_{UV}:=\A_{UV}-\A_\text{Regge}^\text{tree} \,.
\end{equation}
Everything is now manifestly finite for $t\to0^-$. In addition, the second line of Eq.~\eqref{eq:dispersion3} is positive in this limit by unitarity and in the third line the terms in parentheses are negligible. The final term is Planck-suppressed, but is negative. In cases where $f_\text{tree}(t)=\alpha^\prime f_\text{tree}(\alpha^\prime t)$,
the last term can be estimated as
\begin{align}\label{eq:negative_contribution}
    -\frac{2f_\text{tree}^\prime(0)}{\Mpl^2\alpha^\prime}
    =-\mathcal{O}\left(\frac{\alpha^\prime}{\Mpl^2}\right) \,
\end{align}
for $t\to0$. The assumption here indicates that only the string scale appears in $s\to\infty$, which looks reasonable, and this is indeed the case for the Veneziano-Shapiro amplitude in string theory.
To summarize, the approximate gravitational positivity bound at tree level is
\begin{align}\label{eq:approximate}
    \partial_s^2\Delta\A(0,0,0) - \frac{2}{\pi}\int_0^{\Lambda^2}\d{s^\prime}\,\frac{\disc_s\A_\text{IR}(s^\prime,0,u^\prime)}{s^{\prime3}} - \frac{2}{\pi}\int_0^{\Lambda^2}\d{u^\prime}\,\frac{\disc_u\A_\text{IR}(s^\prime,0,u^\prime)}{u^{\prime3}}\geq
    -\mathcal{O}\left(\frac{\alpha^\prime}{\Mpl^2}\right).
\end{align}
In Sec.~\ref{sec:deformed} we will see two examples of Reggeized tree-level amplitudes for which $f_\text{tree}^\prime(0)$ is calculable and allows for the slight negativity argued for above. More generally, we argue in Sec.~\ref{sec:oneloop} that loop effects will generate such a term.


\section{Modified amplitudes}
\label{sec:deformed}

Before discussing one-loop amplitudes, we discuss two explicit examples of Reggeized tree-level amplitudes where the negative $\mathcal{O}(t^0s^2)$ contribution appears from the Regge states.
Explicitly, we discuss the deformed Virasoro-Shapiro amplitude identified recently in~\cite{Arkani-Hamed:2020blm} and the Coon amplitude~\cite{Coon:1969yw,Figueroa:2022onw,Geiser:2022exp,Geiser:2022icl,Chakravarty:2022vrp,Maldacena:2022ckr}, which is a $q$-deformation of the Veneziano amplitude (recently generalized in~\cite{Cheung:2022mkw}).

\subsection{Virasoro-Shapiro amplitude}

As a typical example of Regge-behaved amplitudes in the asymptotic limit, let us consider the Virasoro-Shapiro amplitude which is identified with a 2-to-2 scattering of massless (gauge neutral) bosons in closed string theory. In Ref.~\cite{Tokuda:2020mlf}, this amplitude has been studied in order to check the cancellation of the $t$-channel graviton exchange and the violation of strict positivity. 
In this subsection, we will discuss how the Virasoro-Shapiro amplitude gives the $\mathcal{O}(t^{-1})$ term, but a $\mathcal{O}(t^0)$ term does not appear (see Ref.~\cite{Tokuda:2020mlf} for more detail). Next, we will show that the deformation of the Virasoro-Shapiro amplitude considered in Ref.~\cite{Arkani-Hamed:2020blm}, which maintains the power law for $s$ in the Regge limit, leads to a negative $\mathcal{O}(t^0)$ term.

The Virasoro-Shapiro amplitude is given by the following form:
\begin{align}\label{Eq: Virasoto-Shapiro}
    \A(s,t)=- K(s,t)  \left. \frac{\Gamma(-\alpha(s))\Gamma(-\alpha(t))\Gamma(-\alpha(u))}{\Gamma(1+\alpha(s))\Gamma(1+\alpha(t))\Gamma(1+\alpha(u))}\right| _{u=-s-t}\,,
\end{align}
where $K(s,t)$ is a kinematic factor and $\alpha(x)=\alpha^\prime x/4$. For simplicity, we restrict our attention to the following form of $K(s,t)$:
\begin{align}
    K(s,t) = P \left( s^2 t^2 + t^2 u^2 + s^2 u^2\right)\,,
\end{align}
where $P$ is a positive constant. 
By using the Stirling formula, $\Gamma(z) \sim \sqrt{2\pi}\, z^{z-\frac{1}{2}} e^{-z}$, one can easily see that the asymptotic behavior of $\A(s,t)$ is
\begin{align}\label{Eq: Regge behavior V-S}
    \A(s,t)\sim F_{\mathrm{VS}}(t)\alpha(s)^{2\alpha(t)+2}\,,
\end{align}
where
\begin{align}
    F_{\mathrm{VS}}(t)=P\left(\frac{4}{\alpha^\prime }\right)^{4}   e^{\pi i \alpha(t)}  \frac{\Gamma(-\alpha(t))}{\Gamma(1+\alpha(t))}\,.
\end{align}
Note that $F_{\mathrm{VS}}(t)$ has poles on the real axis at $\alpha(t)=n$ with $n\geq 0$.
The residue in the $t$-plane is 
\begin{equation}\label{Eq: residue V-S}
\begin{aligned}
    \res_{\alpha(t)=n}\A(s,t)&=- K\left(s, \frac{4n}{\alpha^\prime}\right)  \left( \frac{\Gamma(n+\alpha(s))}{n!\Gamma(1+\alpha(s)) }\right)^2
    =-K\left(s, \frac{4n}{\alpha^\prime}\right) \frac{1}{\left( n!\right)^2 } \prod_{j=1}^{n-1}\left( \alpha(s)+j \right)^2\,.
\end{aligned}
\end{equation}
It is clear that the residue at $\alpha(t)=n$ is a polynomial in $s$ of order $2n+2$, and, in particular at $t=0$, the residue is
\begin{align}\label{Eq: residue t=0}
    -\frac{64 P s^2}{\alpha^{\prime3}} \,.
\end{align}
The pole expansion in the $t$- and $u$-channel is thus given by 
\begin{align}
    \A(s,t) = -\frac{4}{\alpha^\prime}\sum_{n=0}^{\infty} 
    \left( \frac{\left(\alpha(s)+1 \right)_{n-1} }{n!}\right)^2
    \left[
    \frac{K\left(s, 4n\alpha^{\prime-1}\right)}{t-4n\alpha^{\prime-1} } 
    - 
    \frac{K\left(s, -s-4n\alpha^{\prime-1}\right)}{t+s+4n\alpha^{\prime-1}}
    \right]\,,
\label{eq:pole_expansion}\end{align}
where we have used the Pochhammer symbol defined as
\begin{align}
    (x)_{m}:=\frac{\Gamma(x+m)}{\Gamma(x)}=\prod_{j=0}^{m-1} (x+j)\,.
\end{align}
For $s\to \pm i\infty$, in particular, the leading contribution from $t$-channel poles is
\begin{align}
    \A(s,t) \sim - P \left( \frac{4}{\alpha^\prime}\right)^5\sum_{n=0}^{\infty} \frac{1}{(n!)^2}\frac{\alpha(s)^{2n+2}}{t-4n\alpha^{\prime-1}} \,.
    \label{eq: pole_expansion_large_s}
\end{align}
Note that this expression as an infinite sum coincides with the Regge behavior \eqref{Eq: Regge behavior V-S} with $t$-channel poles expanded since $F_{\mathrm{VS}}(t)$ can be expressed as
\begin{align}
    F_{\mathrm{VS}}(t)=-P\left( \frac{4}{\alpha^\prime}\right)^5 \sum_{n=0}^{\infty}\frac{1}{(n!)^2}\frac{1}{t-4 n\alpha^{\prime-1}}\,.
\end{align}
By using \eqref{Eq: residue V-S} with $s\leftrightarrow t$ crossing symmetry, we find
\begin{equation} 
\begin{aligned}
    \disc_s \A(s,t) &= -\sum_{n=0}^{\infty}\delta(\alpha(s)-n)\res_{\alpha(s)=n}\A(s,t)\\
    &=  \frac{4}{\alpha^\prime}\sum_{n=0}^{\infty} K\left(\frac{4n}{\alpha^\prime},t\right)
    \left( \frac{\left(\alpha(t)+1 \right)_{n-1} }{n!}\right)^2\delta(s-4n\alpha^{\prime-1}) \,,
\end{aligned}
\end{equation}
and then the integral that we would like to evaluate is 
\begin{equation}\label{Eq: twice derivative V-S}
\begin{aligned}
    \int_{\Lambda^{2}}^{\infty}\d{s^\prime}\,\frac{\disc_s \A(s^\prime,t)}{s^{\prime3}}&=\frac{16  P}{\alpha^{\prime2}}\sum_{n=1}^{\infty}\frac{1}{n}\left( \frac{\left( \alpha(t)+1\right)_{n-1} }{(n-1)!} \right)^{2} + \mathcal{O}(t)\\
    &=-\frac{32 P}{\alpha^{\prime3}t} + \mathcal{O}(t) \,,
\end{aligned}
\end{equation}
where we assume $1/\alpha^\prime>\Lambda^2$. 
The first term on the right-hand side must ensure the cancellation of the graviton $t$-channel pole, and hence we find
\begin{align}\label{Eq: graviton coupling}
    \frac{32 P}{\alpha^{\prime3}}\sim \Mpl^{-2} \,.
\end{align}
Note that we do not obtain any $\mathcal{O}(t^0)$ contributions from the Virasoro-Shapiro amplitude \eqref{Eq: Virasoto-Shapiro}.

Let us next consider the following deformation:
\begin{align}
    \A(s,t)\to \A(s,t)+  \delta\A(s,t) \,,
\end{align}
where $\delta\A$ is defined as
\begin{align}\label{Eq: deformation V-S}
    \delta\A(s,t) = -\epsilon K(s,t)  \frac{\Gamma(1-\alpha(s))\Gamma(1-\alpha(t)\Gamma(1 -\alpha(u))}{\Gamma(2+\alpha(s))\Gamma(2+\alpha(t))\Gamma(2+\alpha(u))} \,.
\end{align}
The positivity of residues requires that the deformation parameter should be bounded as $0\leq \epsilon \leq1$ \cite{Arkani-Hamed:2020blm}. Note that $\delta\A$ can be expressed in terms of $\A$ as follows:
\begin{align}
    \delta\A(s,t) =\epsilon \frac{\alpha(s)\alpha(t)\alpha(u)}{(1+\alpha(s))(1+\alpha(t))(1+\alpha(u))} \A(s,t) \,.
\end{align}
It is hence straightforward to see the Regge behavior of $\delta\A(s,t)$ is
\begin{align}\label{Eq: Regge behavior deltaA}
    \delta\A(s,t) \sim \tilde{F}_{\mathrm{VS}} (t)\alpha(s)^{2\alpha(t)+2} \,,
\end{align}
where 
\begin{align}
    \tilde{F}_{\mathrm{VS}} (t)= \epsilon\frac{\alpha(t)}{1+\alpha(t)} F_{\mathrm{VS}} (t)
    =\epsilon P \left( \frac{4}{\alpha^\prime} \right)^4 e^{\pi i \alpha(t)}\frac{\Gamma(1-\alpha(t))}{\Gamma(2+\alpha(t))} \,.
\end{align}
Note that $\delta\A$ has the same soft behavior as $\A$ in the high energy region with $t<0$, but the first $t$-channel pole appears at $\alpha(t)=1$ rather than at $t=0$, which corresponds to the exchange of a massive particle with spin $4$.
The residue of $\delta\A$ is
\begin{align}\label{Eq: residue deformed V-S}
    \res_{\alpha(s)=n}\delta\A(s,t)
    =  K\left( \frac{4n}{\alpha^\prime},t\right) \frac{\left( \alpha(t) \right)_{n-1} \left( \alpha(t)+2 \right)_{n-1}}{(n-1)!(n+1)!} \,,
\end{align}
and the integral is evaluated as
\begin{equation}
\begin{aligned}
    \int_{\Lambda^{2}}^{\infty}\d{s^\prime}\,\frac{\disc_s \delta \A(s^\prime,t)}{s^{\prime3}}&=-\frac{16 \epsilon P }{\alpha^{\prime2}}\sum_{n=0}^{\infty}\left( n+1 \right) \frac{\left(\alpha(t) \right)_{n}\left(\alpha(t)+2 \right)_{n} }{n! (n+2)!}+ \mathcal{O}(t)\\
    &=-\frac{8\epsilon P}{\alpha^{\prime2}} {}_{2}F_{3} \left( 2,\alpha(t),\alpha(t)+2; 1,3; 1\right) + \mathcal{O}(t)\\
    &=-\frac{8\epsilon P }{\alpha^{\prime2}} + \mathcal{O}(t) \,.
\end{aligned}
\end{equation}
We can thus obtain the $\mathcal{O}(t^0)$ negative term by deforming the Virasoro-Shapiro amplitude by $\delta\A$ defined in~\eqref{Eq: deformation V-S}.
One can also obtain the same result by using~\eqref{Eq: twice derivative V-S} and~\eqref{Eq: Regge behavior deltaA}:
\begin{equation}
\begin{aligned}
    \int_{\Lambda^{2}}^{\infty}\d{s^\prime}\,\frac{\disc_s \A(s^\prime,t)}{s^{\prime3}} &\sim 
    \epsilon \frac{\alpha(t)}{1+\alpha(t)} \int_{\Lambda^{2}}^{\infty}\d{s^\prime}\,\frac{\disc_s \A(s^\prime,t)}{s^{\prime3}}\\
    &=-\frac{8\epsilon P }{\alpha^{\prime2}} + \mathcal{O}(t)\,.
\end{aligned}
\end{equation}
From~\eqref{Eq: graviton coupling} and the bound on $\epsilon$, we find the bound on this negative contribution:
\begin{align}
    \frac{8\epsilon P }{\alpha^{\prime2}}\lesssim \frac{\alpha^\prime}{\Mpl^{2}} \,.
\end{align}
This bound on the negativity is consistent with the recent paper~\cite{Noumi:2022wwf}, in which constraints on Regge amplitudes are considered by using the finite-energy sum rules and IR part of the null constraints.

\subsection{Coon amplitude}

The Coon amplitude~\cite{Coon:1969yw} is a $q$-deformation of the Veneziano amplitude which enjoys many of the same UV properties~\cite{Figueroa:2022onw,Geiser:2022exp,Geiser:2022icl,Chakravarty:2022vrp} but has some unusual features, most notably a spectrum with nonlinear Regge trajectories and an accumulation point below which there are infinitely many states. Despite some similarities with open string scattering in AdS~\cite{Maldacena:2022ckr}, to date the Coon amplitude has no known worldsheet origin.
Indeed, the unitarity of the Coon amplitude has recently been called into question~\cite{Jepsen:2023sia}.
Nevertheless, we take it as a well-studied amplitude and use it as a building block in a crossing-symmetric amplitude featuring both Regge behavior and a $t$-channel pole characteristic of graviton exchange.

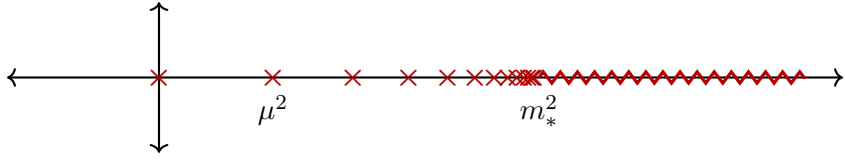
\begin{figure}[t]
    \centering
    \definecolor{myRed}{rgb}{0.7,0,0}
    \begin{tikzpicture}
        \draw[thick, <->] (-2,0) -- (9,0);
        \draw[thick, <->] (0,-1) -- (0,1);
        \foreach \x in {0,1,1.7,2.19,2.53,2.77,2.94,3.06,3.14,3.2,3.24,3.27,3.29}{
            \node[myRed] at (1.5*\x,0) {$\bm{\times}$};
        };
        
        \draw[myRed, very thick, decorate, decoration={zigzag, amplitude=0.75mm, segment length=2.25mm}] (5,0) -- (8.5,0);

        \node[below] at (1.5,-0.1) {$\mu^2$};
        \node[below] at (5,-0.1) {$m_\ast^2$};
    \end{tikzpicture}
    \caption{Schematic spectrum for the Coon amplitude with $\delta=0$. Massive states begin at $m_1^2=\mu^2$ and have an accumulation point at $m_\ast^2=\frac{\mu^2}{1-q}$ where a branch cut begins.}
    \label{fig:Coon_spectrum}
\end{figure}

The Coon amplitude has the following product representation,\footnote{The Veneziano amplitude is recovered in the limit $q\to1^-$.}
\begin{equation}
    \A_q(s,t) = g^2(1-q)\exp\left(\frac{\log{\sigma}\log{\tau}}{\log{q}}\right)\prod_{n=0}^\infty \frac{(\sigma\tau - q^n)(1-q^{n+1})}{(\sigma-q^n)(\tau-q^n)} \,,
\end{equation}
where $g$ is some coupling constant, $q\in(0,1)$ and $\sigma,\tau$ are related to the usual Mandelstam variables as
\begin{equation}
    \sigma = 1 + (q-1)\Big(\frac{s}{\mu^2}-\delta\Big) \,, \qquad \tau = 1 + (q-1)\Big(\frac{t}{\mu^2}-\delta\Big) \,.
\end{equation}
The spectrum is quickly identified to be
\begin{equation}
    m_n^2 = \mu^2(\delta+[n]_q) \,, \qquad [n]_q := \frac{1-q^n}{1-q} \,,
\end{equation}
with an accumulation point for large $n$:
\begin{equation}
    m_n^2 \quad\xrightarrow{n\to\infty}\quad m_\ast^2 := \mu^2\left(\delta + \frac{1}{1-q}\right) \,.
\end{equation}
There is a cut from the $\log{\sigma}$ factor which extends from the branch point at $s=m_\ast^2$ out to infinity: see Fig.~\ref{fig:Coon_spectrum} for a sketch. Going forward we take $\delta=0$ so that the spectrum contains massless states.

At low energies, $s,t,u\ll\mu^2$, one finds
\begin{equation}
    \A_q(s,t) \approx -g^2\mu^2\left(\frac{1}{s}+\frac{1}{t}\right) \,,
\end{equation}
so that the crossing-symmetric combination
\begin{equation}\label{eq:Coon_crossing}
    \tilde{\A}_q(s,t) = (s^2+t^2+u^2)\big[\A_q(s,t)+\A_q(t,u)+\A_q(u,s)\big]
\end{equation}
at low energies features the $t$-channel pole characteristic of tree-level graviton exchange:
\begin{equation}\label{eq:Coon_lowenergy}
    \tilde{\A}_q(s,t) \approx g^2\mu^2\,\frac{(s^2+t^2+u^2)^2}{stu} \quad\xrightarrow{t\to0}\quad -4g^2\mu^2\,\frac{s^2}{t} \,.
\end{equation}
In the Regge limit the Coon amplitude takes the form
\begin{equation}
    \A_q(s,t) \quad\longrightarrow\quad g^2(1-q)\sigma^{\frac{\log{\tau}}{\log{q}}}\prod_{n=0}^\infty \frac{1-q^{n+1}}{1-\frac{q^n}{\tau}} \approx f_q(t)\left(-\frac{s}{m_\ast^2}\right)^{\frac{\log{\tau}}{\log{q}}}
\end{equation}
and so
\begin{equation}
    \tilde{\A}_q(s,t) \quad\longrightarrow\quad \tilde{f}_q(t)\left(-\frac{s}{m_\ast^2}\right)^{2+\frac{\log{\tau}}{\log{q}}} \,,
\end{equation}
which marginally satisfies Eq.~\eqref{eq:s2bounded} for $t<0$ ($\tau>1$) since $\log{q}$ is negative. The contour integral around the positive real-$s$ axis splits into a sum over poles and an integral along the branch cut:
\begin{equation}
    \int_{\mu^2}^\infty\d{s^\prime}\,\frac{\disc_s\A_q(s^\prime,t)}{s^{\prime3}} = -\pi\sum_{k=1}^\infty\frac{1}{m_k^6}\res_{s=m_k^2}\A_q(s,t) + \int_{m_\ast^2}^\infty\d{s^\prime}\,\frac{\disc_s\A_q(s^\prime,t)}{s^{\prime3}} \,.
\end{equation}
The residues are polynomial in $t$,
\begin{equation}
    \res_{s=m_k^2}\A_q(s,t) = g^2\mu^2q^k\prod_{n=0}^{k-1}\frac{\tau-q^{n-k}}{1-q^{n-k}} \quad\xrightarrow{t\to0}\quad g^2\mu^2q^k\,,
\end{equation}
and the sum over $k$ gives a finite contribution for $t\to0$. In contrast, the integral over the branch cut diverges for $t\to0$ and cancels the $t$-channel pole. The range $s>m_\ast^2$ corresponds to $\sigma<0$, so we have 
\begin{align}
    \disc_s\A_q(s,t) &= -g^2(1-q)(-\sigma)^{\frac{\log\tau}{\log q}}\sin{\left(\pi\,\frac{\log\tau}{\log q}\right)}\prod_{n=0}^\infty \frac{(\sigma\tau - q^n)(1-q^{n+1})}{(\sigma-q^n)(\tau-q^n)} \notag\\
    &= g^2(1-q)(-\sigma)^{\frac{\log\tau}{\log q}}\sin{\left(\pi\,\frac{\log\tau}{\log q}\right)}\frac{(\sigma\tau - 1)}{(\sigma-1)\frac{t}{\mu^2}}\prod_{n=1}^\infty \frac{(\sigma\tau - q^n)(1-q^{n+1})}{(\sigma-q^n)(\tau-q^n)} \notag\\
    &= -\pi g^2(-\sigma)^{-\frac{1-q}{\log q}\frac{t}{\mu^2}}\frac{1-q}{\log q}\big[1 + \mathcal{O}(t)\big]
\end{align}
and thus
\begin{equation}
    \frac{\disc_s\tilde{\A}_q(s^\prime,t)}{(s^\prime)^3} \approx -2\pi g^2\mu^2 \left(\frac{s^\prime}{\mu^2}\right)^{-1-\frac{1-q}{\log{q}}\frac{t}{\mu^2}}\frac{1-q}{\log q}\big[1 + \mathcal{O}(t)\big]
\end{equation}
gives
\begin{equation}
\begin{aligned}
    \int_{m_\ast^2}^\infty\d{s^\prime}\,\frac{\disc_s\tilde{\A}_q(s^\prime,t)}{(s^\prime)^3} &\approx -2\pi g^2\mu^{-2} \frac{1-q}{\log q}\big[1 + \mathcal{O}(t)\big]\int_{m_\ast^2}^\infty\d{s^\prime}\,\left(\frac{s^\prime}{\mu^2}\right)^{-1-\frac{1-q}{\log{q}}\frac{t}{\mu^2}}\\
    &= -\frac{2\pi g^2\mu^2}{t} \big[1 + \mathcal{O}(t)\big]
\end{aligned}
\end{equation}
This leading $1/t$ behavior exactly cancels the $t$-channel pole of Eq.~\eqref{eq:Coon_lowenergy} in the dispersion relations. Here the details of the cancellation are somewhat different, however, since the sum over poles does not generate a $1/t$ contribution despite there being an infinite number of states. Knowing the contributions from poles and the branch cut must exactly compensate for the massless state, the $\mathcal{O}(t^0)$ can be shown to be nonzero,
\begin{equation}
    4g^2\mu^2\left( -\frac{1}{t} + \frac{1-q}{\mu^2} + \mathcal{O}(t) \right) \,,
\end{equation}
providing another example where a Reggeized amplitude leaves a finite contribution after canceling the $1/t$ divergence.
We emphasize, however, that although $\tilde{\A}_q(s,t)$ as introduced in Eq.~\eqref{eq:Coon_crossing} has the features we argue for generally, it is not known if this amplitude arises in a fully-consistent gravitational theory.


\section{Positivity bounds at one loop}
\label{sec:oneloop}

In this section, we first see how the positivity bound at the one-loop level leads to an apparent puzzle that the cut-off scale of the theory is extremely low, under the assumption that the graviton $t$-channel pole is neglected.
Then, we propose how the Reggeization extends to one-loop level.

\subsection{Small cut-off scale}\label{Sec:small_cutoff}
Suppose that the approximate gravitational positivity bound~\eqref{eq:approximate} at tree-level is correct even at the one-loop level.
Then, Ref.~\cite{Alberte:2020bdz} found that the left-hand side of \eqref{eq:approximate} is computed as
\begin{align}
    &\partial_s^2\Delta\A(0,0,0) - \frac{2}{\pi}\int_0^{\Lambda^2}\d{s^\prime}\,\frac{\disc_s\A_\text{IR}(s^\prime,0,u^\prime)}{s^{\prime3}} - \frac{2}{\pi}\int_0^{\Lambda^2}\d{u^\prime}\,\frac{\disc_u\A_\text{IR}(s^\prime,0,u^\prime)}{u^{\prime3}}
    \nonumber\\&\approx
    \frac{e^4}{4\pi^2\Lambda^4}
    -\frac{e^2}{180\pi^2m^2\Mpl^2} \,.
\label{eq:EFT_amplitude}\end{align}
Note that the second term in the second line corresponds to a gravitational diagram while the first term corresponds to a non-gravitational diagram.
Then, using \eqref{eq:approximate}, we obtain
\begin{align}
    \frac{e^4}{4\pi^2\Lambda^4}
    -\frac{e^2}{180\pi^2m^2\Mpl^2}
    \geq
    -\mathcal{O}\left(\frac{\alpha^\prime}{\Mpl^2}\right),
\end{align}
which leads to the upper bound on $\Lambda$:
\begin{equation}
\label{eq:positivitybounds}
    \Lambda \lesssim \sqrt{em\Mpl} \,,
    \qquad\quad \text{for $m^2\ll e^2\alpha^{\prime-1}$} \,.
\end{equation}
Similar results are reported for the Standard Model~\cite{Aoki:2021ckh} and the dark photon model~\cite{Noumi:2022zht}.
In particular, using this argument, it is argued that almost all parameter regions of the dark photon model are excluded~\cite{Noumi:2022zht}.

\begin{figure}[t]
    \begin{center}
        \includegraphics[width=80mm, height=60mm]{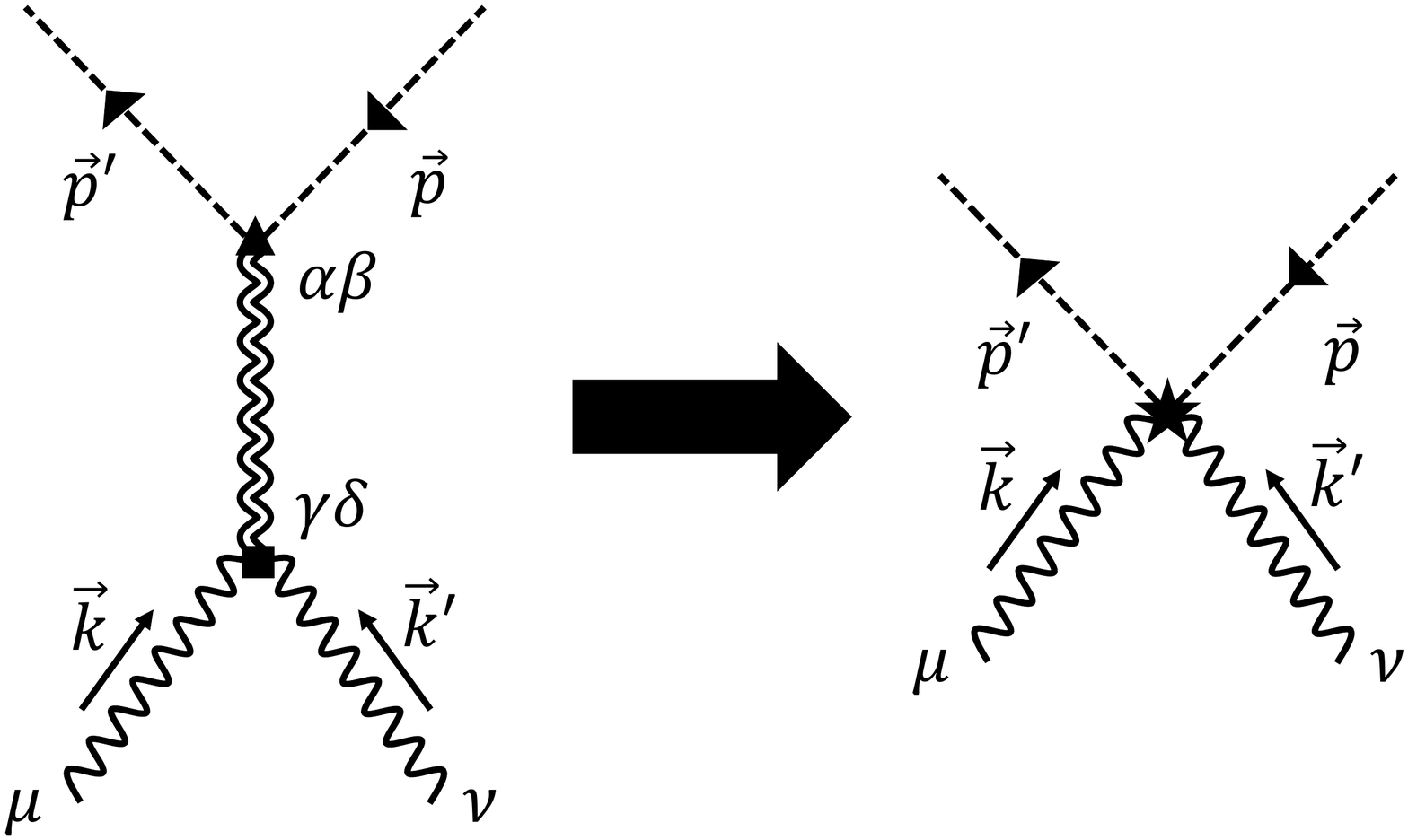}
        \caption{A star vertex in the right diagram is a new vertex from Eq.~\eqref{eq:action41}. We consider new diagrams obtained by replacing the left diagram with the right one.}
        \label{fig41}
    \end{center}
\end{figure}

Note that even if a higher-dimensional term is added to quadratic action~\eqref{eq:scalarQEDaction}, there is no change in the positivity bound~\eqref{eq:positivitybounds}.
For example, we can consider the additional term with mass dimension six, 
\begin{equation}
    \label{eq:action41}
    S_6 = \int\d[4]{x}\,\sqrt{-g}\frac{c}{\Lambda^2}\phi^\dagger\phi F_{\mu\nu}F^{\mu\nu}\,,
\end{equation}
where $c$ is a constant.
A new vertex (Fig.~\ref{fig41}) and diagrams arise from Eq.~\eqref{eq:action41}, but we can confirm that their contributions to the left-hand side of~\eqref{eq:approximate} cancel out.

Originally, the bound~\eqref{eq:positivitybounds} is derived by assuming that the graviton $t$-channel pole can be discarded, which is almost equivalent to the approximately gravitational positivity bound~\eqref{eq:approximate} at tree-level.
After that, it turns out that in the case of graviton-photon~\cite{Alberte:2021dnj} and graviton-graviton~\cite{deRham:2022gfe} scatterings, the $t$-channel pole can not be neglected.
This implies that the one-loop Regge amplitude cancels the negative contribution corresponding to the graviton exchange.

In the following, as an independent argument for the importance of the contribution from the Regge states, we provide a counterexample of Eq.~\eqref{eq:positivitybounds} in string theory, assuming that the inequality similar to Eq.~\eqref{eq:positivitybounds} holds in the case of 4d $\mathcal{N}=2$ theory.
Let us take type IIB theory compactified on Calabi-Yau threefold.\footnote{We can also consider the IIA mirror dual.} 
The number of supersymmetry is reduced from $32$ to $8$, and then the resultant 4d theory has $\mathcal{N}=2$ supersymmetry. The theory has a $h^{2,1}$ number of $U(1)$ gauge symmetries corresponding to the dimensional reduction of 10d Ramond-Ramond 4-form field, where $h^{2,1}$ is the Hodge number of the Calabi-Yau threefold.
On top of that, there are $h^{1,1}+1$ number of neutral hypermultiplets, which are not important in our context. 
A part of the moduli space is parameterized by the VEV of the scalar partner of $U(1)$ gauge field.
In general, this moduli space has a conifold singularity, where the $D3$-brane wrapping on the internal cycle becomes massless~\cite{Strominger:1995cz}.
In the 4d language, this massless particle is a 4d $\mathcal{N}=2$ hypermultiplet charged under $U(1)$, since $D3$-brane is charged under Ramond-Ramond 4-form in 10d.
By choosing the VEV slightly away from the conifold point, we have an electron (charged hypermultiplet) with an arbitrarily small mass. In contrast, the mass of the other heavy states, which is interpreted as the cutoff scale, is finite.\footnote{When the distance to the singularity is finite, only a finite number of the state becomes massless. On the other hand, an infinite number of states become massless for an infinite distance, as dictated by the Distance Conjecture.}
This is a counterexample to the Eq.~\eqref{eq:positivitybounds} because the right-hand side of the inequality becomes arbitrarily small.
Therefore, we arrive at the conclusion that the Regge amplitude cancels the negative contribution even in the case of photon-photon scattering.
This counterexample also shows that the negative contribution would not be canceled by modifying the Regge slope $2+\alpha^\prime t$, but by the modification of the prefactor $f(t)$ (cf. Eq.~\eqref{eq:Regge_limit}) at the loop level.
This is because the tower of higher spin states does not become massless at the conifold point.
This is a novel point compared with the argument based on sum rules.

\subsection{Reggeization at the loop level}
\label{sec:reggeloop}

In the previous section, we first argue that the bound~\eqref{eq:positivitybounds} follows from the gravitational positivity bound~\eqref{eq:approximate}.
Then, we argue that the bound contradicts with the well-established string theory vacua.
This indicates that the derivation of Eq.~\eqref{eq:positivitybounds} in Sec.~\ref{Sec:Cancellation} should be revisited.

In the following, we argue that the Regge amplitude at the loop level changes the argument leading to Eq.~\eqref{eq:positivitybounds}.
In Eq.~\eqref{eq:Regge_limit}, we have used $f\approx f_\text{tree}$. Instead, if we keep the quantum correction $f=f_\text{tree}+f_\text{loop}$, then the term 
\begin{align}
    -\frac{2f_\text{loop}^\prime(t)}{\Mpl^2\alpha^\prime}
\label{eq:loop_f}\end{align}
is added to the right-hand side of Eq.~\eqref{eq:dispersion3}.
By comparing this with Eq.~\eqref{eq:EFT_amplitude}, we observe that if the term~\eqref{eq:loop_f} cancels the negative contribution corresponding to the graviton exchange, then it helps with positivity. 
This is a resolution of the contradiction in the previous section.
Naively, the cancellation between Eq.~\eqref{eq:loop_f} and $-e^2/(180\pi^2m^2\Mpl^2)$ is unlikely because the UV Regge amplitude should ``know" the IR mass scale $m$.
However, here we propose the generalization of the Regge amplitude to the one-loop level, which realizes the cancellation naturally.

At the tree level, we have assumed the Reggeization
\begin{align}
    \A\to F_\mathrm{tree}(t)\frac{s^{2+\alpha^\prime t}}{\Mpl^2 M_*^{2\alpha^\prime t}}
\end{align}
for fixed $t$ and large $s$.
Here $F_\mathrm{tree}$ has mass dimension $-2$, and the relation with $f_\text{tree}$ in Eq.~\eqref{eq:Regge_limit} is $F_\mathrm{tree}\sin(\alpha^\prime t)=f_\text{tree}$.
This is expected to happen by the exchange of the higher spin particles.
Given this interpretation, the important requirement to the behavior of $F_\mathrm{tree}(t)$ is that
\begin{itemize}
    \item
    $F_\mathrm{tree}(t)$ must have poles at $t=n\,\alpha^{\prime-1}$ where $n=0,1,\ldots$. Therefore, $F_\mathrm{tree}(t)$ is written as
    \begin{align}
        F_\mathrm{tree}(t) = \frac{f_0}{t}+\frac{f_1}{t-\alpha^{\prime-1}}+\frac{f_2}{t-2\alpha^{\prime-1}}+\cdots \,,
    \end{align}
    where $f_i$ is a constant. 
    Since each residue of the pole must be positive, we obtain $f_i>0$.
    The presence of the pole at $t=0$ is crucial to maintain the positivity bound at the tree-level.
    Because of the poles, the amplitude at $t=n\,\alpha^{\prime-1}$ is dominated by the diagram corresponding to the exchange of the particle of the mass $n\,\alpha^{\prime-1}$. 
\end{itemize}

\noindent
Next, we have to impose an extra requirement at the one-loop level. We denote the amplitude as
\begin{align}
    \A\to\big[F_\mathrm{tree}(t)+F_\mathrm{loop}(t)\big]\frac{s^{2+\alpha^\prime t}}{\Mpl^2 M_*^{2\alpha^\prime t}} \,.
\end{align}
At the one-loop level, contributing diagrams include those with a loop of light particles and tree-level exchange of higher spin particles (see Sec.~\ref{Sec:loop}).
Importantly, these diagrams have a branch cut starting from $t=4m^2$ corresponding to particle production.

\begin{itemize}
    \item At the one-loop level, the branch cut must appear at $t=4m^2$. $F_\mathrm{loop}$ is written as
    \begin{align}
        F_\mathrm{loop}(t)= \tilde{f}_0L_0(t,m^2)+\frac{\tilde{f}_1\alpha^{\prime-1}}{t-\alpha^{\prime-1}}L_1(t,m^2)+\frac{\tilde{f}_2\alpha^{\prime-1}}{t-2\alpha^{\prime-1}}L_2(t,m^2)+\cdots \,,
   \label{Eq:F_loop}\end{align}
    where $L_i(t,m^2)$ is a loop function that contains the branch cut starting from $t=4m^2$ (see Sec.~\ref{Sec:loop} for the detail), and has an expansion
    \begin{align}
        L_i(t,m^2) =\sum_{j=1}^\infty \frac{l_j^{(i)}}{m^2}\left(\frac{t}{m^2}\right)^{j-1} \,,
    \label{Eq:loop_expansion}\end{align}
    up to $\log(m^2)$. Here $l_j$ is a dimensionless constant. 
    On top of the branch cut, there exist poles at $t=k\alpha^\prime$ where $k=1,2,\ldots$.
\end{itemize}

\noindent
Now we discuss the implications of the Regge amplitude on the dispersion relation.
The contribution from Eq.~\eqref{Eq:F_loop} is
\begin{align}
    \int^\infty_{M_*^2} \d{s^\prime}\,\frac{\disc\A(s^\prime,t)}{s^{\prime3}}\sim \frac{\sin(\alpha^\prime t)}{\alpha^\prime t}
    \left(
    \tilde{f}_0L_0(t,m^2)+\frac{\tilde{f}_1\,t}{t-\alpha^{\prime-1}}L_1(t,m^2)+\frac{\tilde{f}_2\,t}{t-2\alpha^{\prime-1}}L_2(t,m^2)+\cdots
    \right).
\end{align}
For $t\to0$, by choosing $\tilde{f}_i=\mathcal{O}(\alpha^\prime)$ appropriately, the inequality~\eqref{eq:dispersion3} contains a finite contribution from the one-loop diagrams with higher-spin spins exchanged. In particular, the positivity bounds for scalar QED can be satisfied without requiring an unusually low cut-off.

\subsection{An argument for the form of \texorpdfstring{$F_\mathrm{loop}$}{Floop}}
\label{Sec:loop}

\begin{figure}[t]
    \begin{center}
        \includegraphics[width=75mm, height=56.25mm]{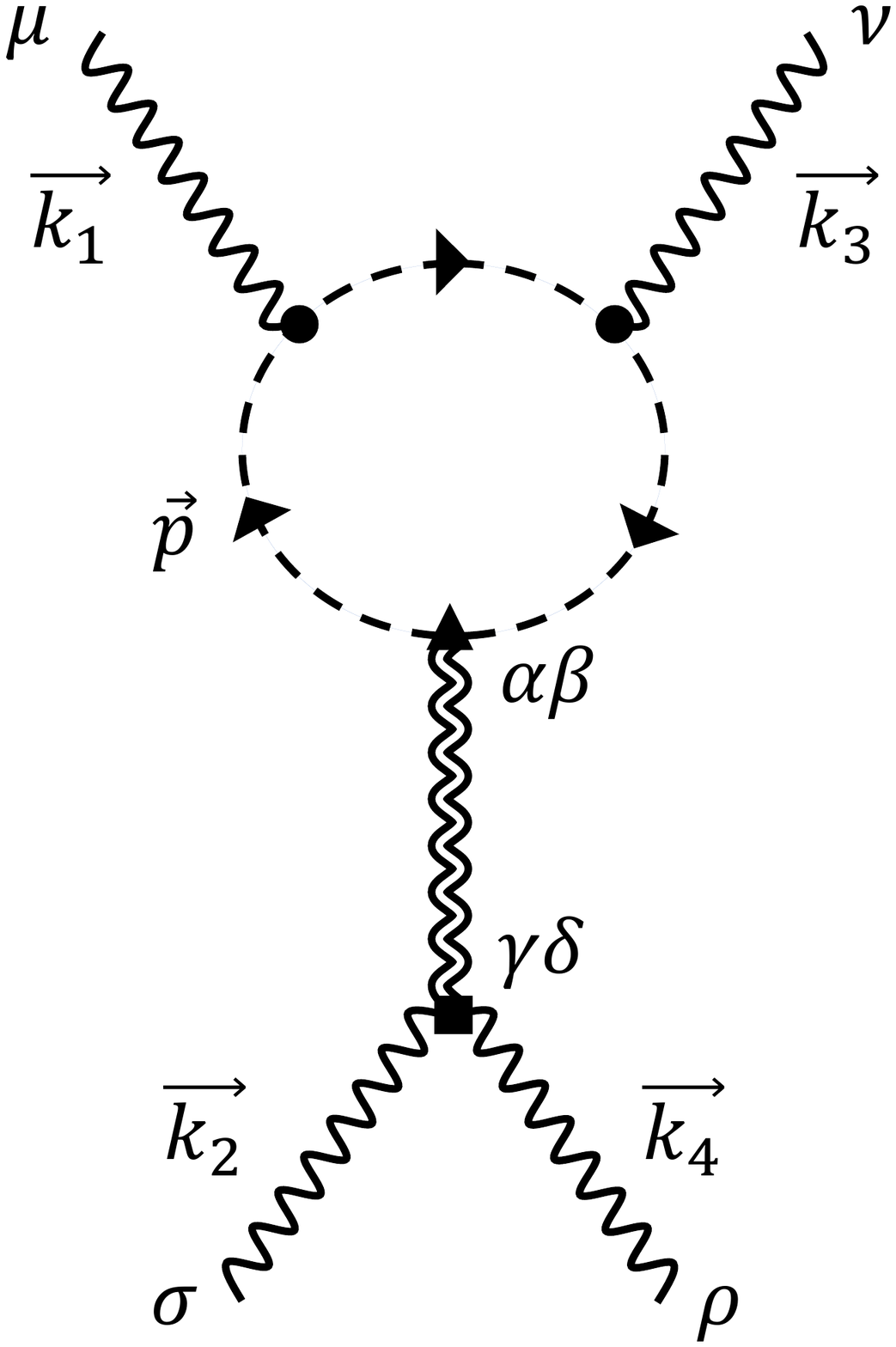}
        \includegraphics[width=75mm, height=56.25mm]{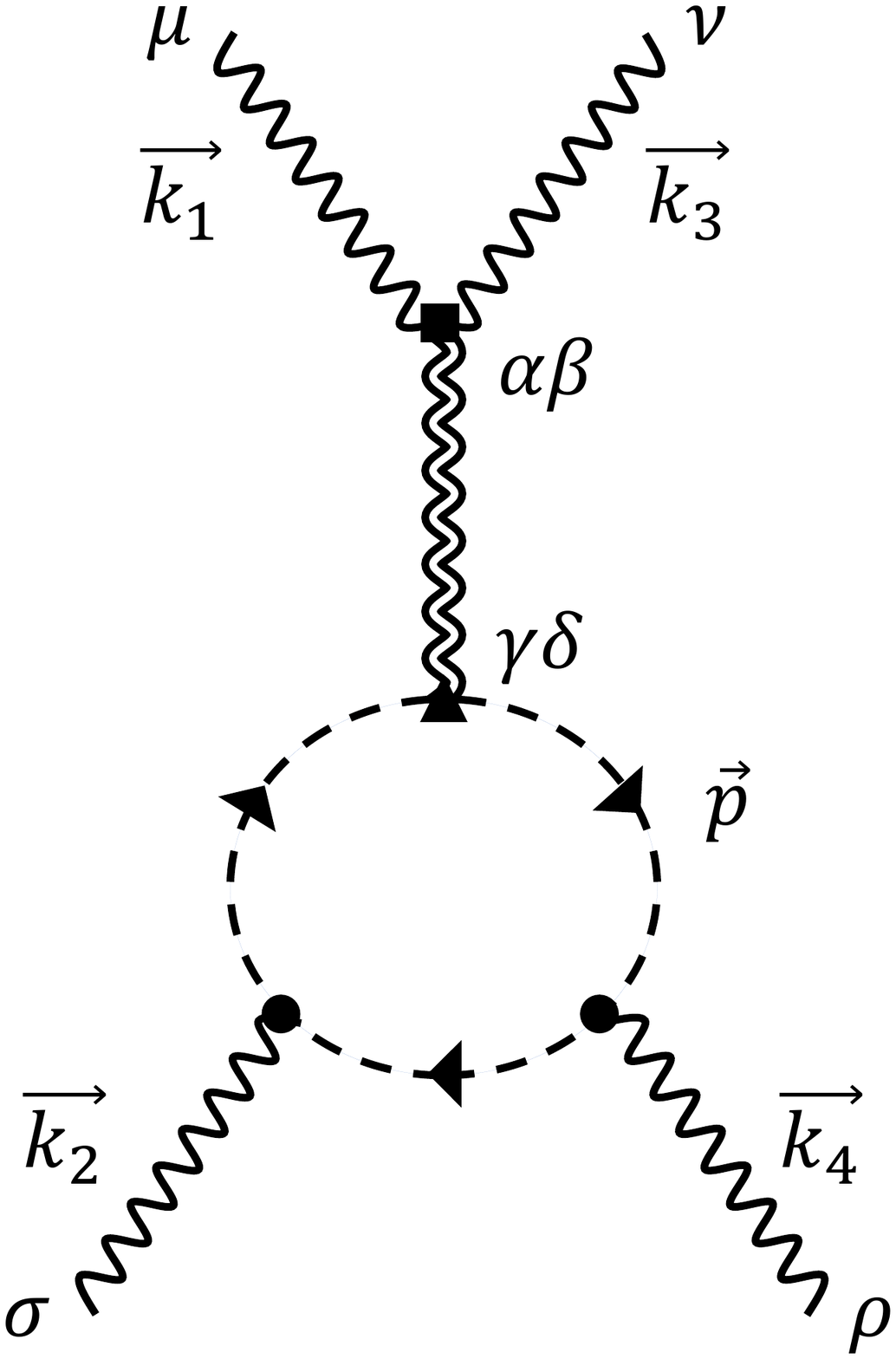}
        \includegraphics[width=75mm, height=56.25mm]{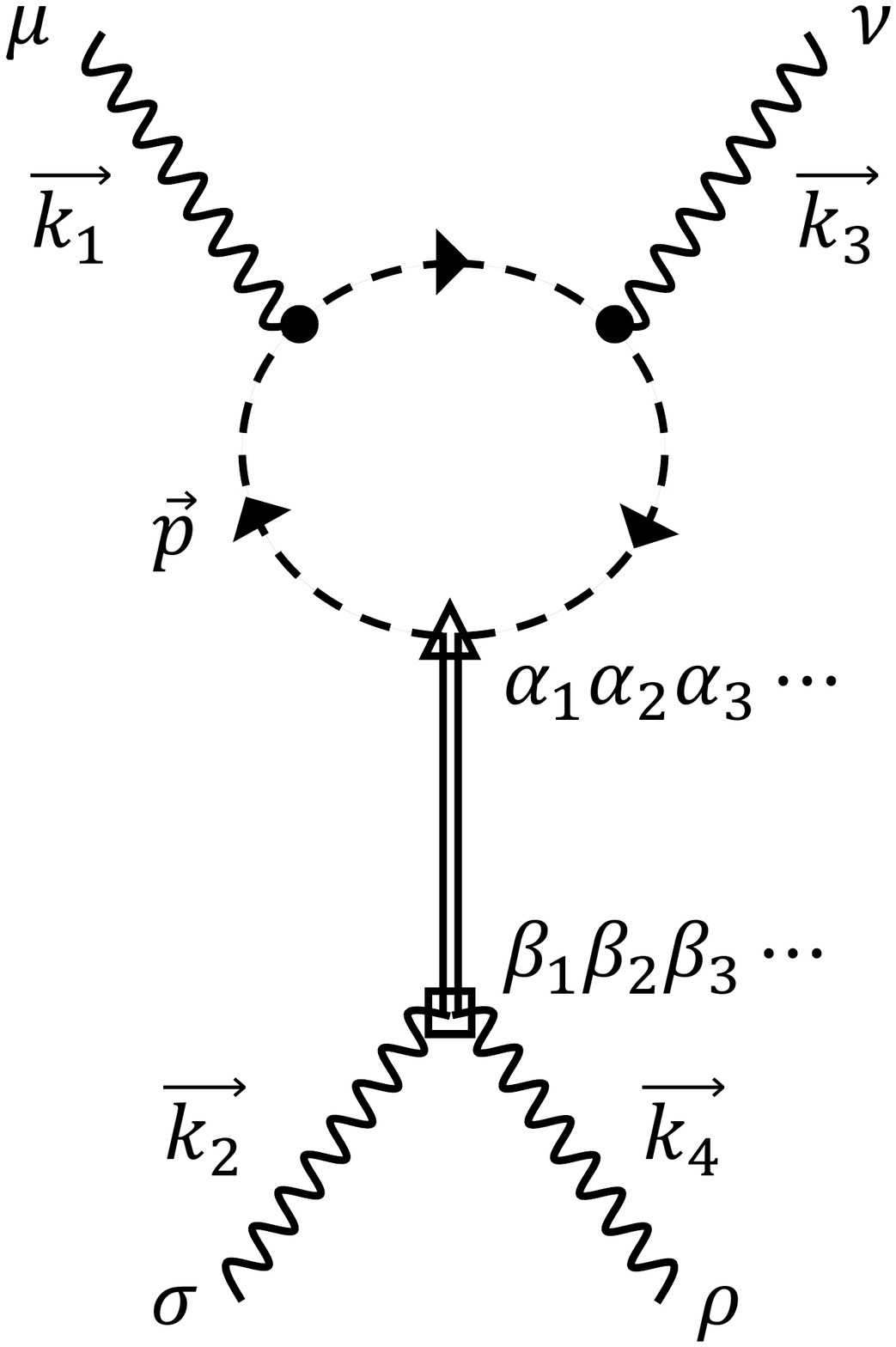}
        \includegraphics[width=75mm, height=56.25mm]{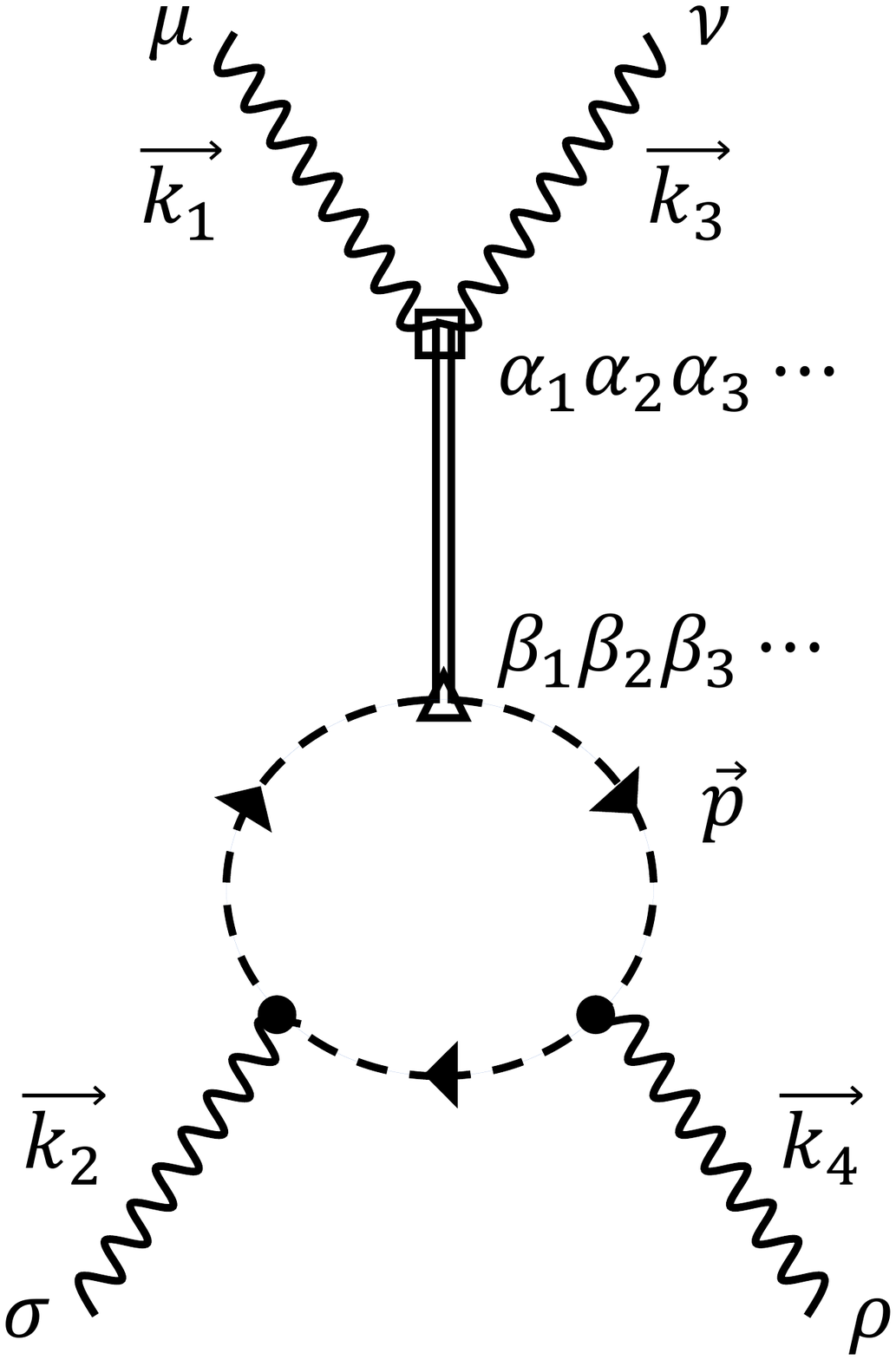}
        \caption{The one-loop diagrams corresponding to the $t$-channel exchange of the graviton and the higher spin particles. 
        The solid, wavy, double solid, and double wavy lines correspond to the scalar, graviton, photon, and higher spin particles, respectively. These produce the branch cut starting from $t=4m^2$.}
        \label{fig:diagram}
    \end{center}
\end{figure}

In this section we provide an explanation for Eq.~\eqref{Eq:F_loop} by estimating the $t$-channel exchange diagrams of Fig.~\ref{fig:diagram}, where the double line is either the graviton or the higher spin particles (see also the caption).
We show that this class of diagrams leads to the contribution where the light mass parameter $m^2$ shows up in the denominator.
In App.~\ref{sec:diagrams}, we argue that the other diagrams do not exhibit this feature by the power counting argument.
The loop integral in Fig.~\ref{fig:diagram} is (see App.~\ref{sec:notation} for the notation)
\begin{equation}
\begin{aligned}
    &\frac{g_1g_2}{M^{2L-2}}\int\frac{\d[4]{p}}{(2\pi)^4}
    \frac{1}{p^2-m^2}\frac{1}{(p+k_1)^2-m^2}\frac{1}{(p+k_1+k_3)^2-m^2}(2p+k_1)^{\mu}(2p+2k_1+k_3)^{\nu}\\
    &\qquad\qquad\times (p^{\rho_1}\cdots p^{\rho_{n_1}})\bigg((p+k_1+k_3)^{\sigma_1}\cdots (p+k_1+k_3)^{\sigma_{n_2}}\bigg) \,,
\end{aligned}
\end{equation}
where the second line comes from the interaction between $\phi$ and higher spin-$L$ particle $\Phi_{\alpha_1\cdots\alpha_L}$,
\begin{align}
    \frac{g_1}{M^{L-1}} \Phi_{\alpha_1\cdots\alpha_L}\partial^{\alpha_1}\cdots\partial^{\alpha_i}\phi\partial^{\alpha_{i+1}}\cdots\partial^{\alpha_L}\phi \,,
\label{Eq:coupling1}\end{align}
and $n_1+n_2=L$ with $M$ and $g_1$ being the mass and coupling of $\Phi_{\alpha_1\cdots\alpha_L}$.
Note that the coupling between the photon and $\Phi_{\alpha_1\cdots\alpha_L}$ is schematically given by
\begin{align}
    \frac{g_2}{M^{L-1}}\Phi_{\beta_1\cdots\beta_L}\partial^{\beta_3\cdots\beta_m}F^{\beta_1\mu}\partial^{\beta_{m+1}\cdots\beta_L}{F^{\beta_2}}_\mu,
\label{Eq:coupling2}\end{align}
where $g_2$ is the coupling constant.
By introducing Feynman parameters, we obtain
\begin{align}
    &\frac{g_1g_2}{M^{2L-2}}\int\frac{\d[4]{l}}{8\pi^4}\int_0^1 \d{y}\int_0^{1-y} \d{x}\,
    \frac{(2l+(1-2x-2y)k_1-2yk_3)^{\mu}(2l+2(1-x-y)k_1+(1-2y)k_3)^{\nu}}{[l^2-m^2+y(1-x-y)t]^3}
    \nonumber\\
    &\times
    \bigg((l-(x+y)k_1-yk_3)^{\rho_1}\cdots (l-(x+y)k_1-yk_3)^{\rho_{n_1}}\bigg)
    \nonumber\\
    &\times\bigg((l+(1-x-y)k_1+(1-y)k_3)^{\sigma_1}\cdots (l+(1-x-y)k_1+(1-y)k_3)^{\sigma_{n_2}}\bigg)
\end{align}
where $l=p+(x+y)k_1+yk_3$.

In the following, we concentrate on the term which contains the light mass parameter $m^2$ in the denominator for $t\to0$. From a power-counting argument, this corresponds to the terms where the numerator does not contain $l$: 
\begin{align}
    &\frac{g_1g_2}{M^{2L-2}}\int\frac{\d[4]{l}}{8\pi^4}\int_0^1 \d{y}\int_0^{1-y} \d{x}\,
    \frac{((1-2x-2y)k_1-2yk_3)^{\mu}(2(1-x-y)k_1+(1-2y)k_3)^{\nu}}{[l^2-m^2+y(1-x-y)t]^3}
    \nonumber\\
    &\times
    \bigg((-(x+y)k_1-yk_3)^{\rho_1}\cdots (-(x+y)k_1-yk_3)^{\rho_{n_1}}\bigg)
    \nonumber\\
    &\times\bigg(((1-x-y)k_1+(1-y)k_3)^{\sigma_1}\cdots ((1-x-y)k_1+(1-y)k_3)^{\sigma_{n_2}}\bigg).
\label{eq:integral2}\end{align}

The structure of the cut becomes clear by integrating over $l$, $x$ and $y$. 
We obtain the following expression:
\begin{align}
    &\int_0^1 \d{y}\int_0^{1-y} \d{x}\int\frac{\d[4]{l}}{8\pi^4}\frac{1}{[l^2-m^2+y(1-x-y)t]^3}
    \nonumber\\
    &\qquad=-i\int_0^1 \d{y}\int_0^{1-y} \d{x}\int\frac{\d[d]{l_E}}{8\pi^4}\frac{1}{[l_E^2+m^2-y(1-x-y)t]^3}
    \nonumber\\
    &\qquad=-i \int_0^1 \d{y}\int_0^{1-y} \d{x}\,\frac{1}{16\pi^2}\frac{1}{m^2-y(1-x-y)t}
    \nonumber\\
    &\qquad=
    -\frac{i}{t}\;\mathrm{Li}_2\left(\frac{2\sqrt{t}}{\sqrt{t}-\sqrt{t-4m^2}}\right)
    -\frac{i}{t}\;\mathrm{Li}_2\left(\frac{2\sqrt{t}}{\sqrt{t}+\sqrt{t-4m^2}}\right) \,,
\label{eq:t-channel_exchange}\end{align}
where $\mathrm{Li}$ is the polylogarithm function.\footnote{Note that Eq.~\eqref{eq:t-channel_exchange} is finite at $t=4m^2$.}
Here for simplicity, we ignore $x$ and $y$ dependence in the numerator, but the structure of the branch cut does not change.
The Eq.~\eqref{eq:t-channel_exchange} has a small-$t$ expansion of the form~\eqref{Eq:loop_expansion}
\begin{equation}
\begin{aligned}
    &-\frac{i}{t}\;\mathrm{Li}_2\left(\frac{2\sqrt{t}}{\sqrt{t}-\sqrt{t-4m^2}}\right)
    -\frac{i}{t}\;\mathrm{Li}_2\left(\frac{2\sqrt{t}}{\sqrt{t}+\sqrt{t-4m^2}}\right)\\
    &=-i\left(
    \frac{1}{2m^2}+\frac{t}{24m^4}+\frac{t^2}{180m^6}+\cdots
    \right) \,.
\end{aligned}
\end{equation}
We write Eq.~\eqref{eq:integral2} in the following schematic way:
\begin{align}
    \eqref{eq:integral2}=(k_1+\# k_3)^\mu(k_1+\# k_3)^\nu 
    \cdots(k_1+\# k_3)^{\sigma_{n_2}} F_{L-2}(t,m^2) \,,
\end{align}
where $F_{L-2}(t,m^2)$ is defined by
\begin{align}
    F_{L-2}(t,m^2)=&(-1)^{n_1}\int\frac{\d[4]{l}}{8\pi^4}\int_0^1 \d{y}\int_0^{1-y} \d{x}\,
    \frac{2(1-x-y)(1-2x-2y)(x+y)^{n_1}(1-x-y)^{n_2}}{[l^2-m^2+y(1-x-y)t]^3}.
\end{align}
The indices $\mu$ and $\nu$ are contracted with the external polarizations $\epsilon_1^\mu$ and $\epsilon_3^\nu$. This gives a factor 
\begin{align}
    (k_1+\# k_3)^\mu(k_1+\# k_3)^\nu \epsilon_{1\mu}\epsilon_{3\nu}
    \sim \frac{tu}{s}\to t \,,
    \quad\text{for $s\to\infty$}
\end{align}
Other indexes are contracted with $k_2, k_4, \epsilon_2$ or $\epsilon_4$. For large $s$, this contribution behaves as $s^L$.
Multiplying the propagator of the higher spin particle, we obtain
\begin{align}
    \text{Figure~\ref{fig:diagram}}\;=
     \frac{g_1g_2}{M^{2L-2}} \frac{s^L t}{t-M^2}F_{L-2}(t,m^2)
\end{align}
for large $s$.
Suppose that the mass $M$ and the spin $L$ of the particle are given by
\begin{equation}
    M^2 = k\alpha^{\prime-1} \,,
    \qquad L=2+k \,,
    \qquad k=0,1,2,\ldots \,,
\end{equation}
as usual in the graviton Regge tower.
Then, the amplitude is
\begin{align}
    \frac{g_1g_2}{M^{2L-2}}\frac{s^{2+k}t}{t-k\alpha^{\prime-1}}F_{k}(t,m^2) \,.
\label{Eq:loop_amplitude}\end{align}
When we take $t\to k\alpha^{\prime-1}$, we expect that the contribution from the exchange of spin $2+k$ particle dominates.
This reproduces Eq.~\eqref{Eq:F_loop}.

We would like to explain an idea why the summation of the $t$-channel diagram leads to the Reggeized amplitude.
At the tree-level, the $t$-channel amplitude is
\begin{align}
    \frac{s^2}{\Mpl^2t}
    +g_2^2\frac{s^3\alpha^{\prime2}}{t-\alpha^{\prime-1}}
    +g_2^2\frac{s^4\alpha^{\prime3}}{t-2\alpha^{\prime-1}}
    +\cdots \,.
\end{align}
After an appropriate analytic continuation, we obtain $s^{2+\alpha^\prime t}/(\Mpl^2t)$ for fixed $t$.\footnote{See Eq.~\eqref{eq:pole_expansion} and App.~\ref{sec:tree_Regge} for the detail.}
Similarly, at the one-loop level, the amplitude is
\begin{align}
    &\left(\frac{s^2}{\Mpl^2}
    +g_1g_2\frac{s^{3}t\,\alpha^{\prime2}}{t-\alpha^{\prime-1}}
    +g_1g_2\frac{s^{4}t\,\alpha^{\prime3}}{t-2\alpha^{\prime-1}}+\cdots
    \right)F(t,m^2)\,,
\end{align}
from Eq.~\eqref{Eq:loop_amplitude}.
Here we have assumed that all $F_k$ are the same order of magnitude, and have defined $F:=F_k$.
Since the expression in the parenthesis is the same form as the tree-level $t$-channel amplitude, it is natural to expect that this is also Reggeized as $s^{2+\alpha^\prime t}/\Mpl^2$ for fixed $t$.
In fact, the formal summation of the $t$-channel diagram leads to
\begin{align}
    \frac{s^2}{\Mpl^2}
    +g_1g_2\sum_{n=1}^\infty\frac{s^{2}t\,\alpha^{\prime}}{t-n\alpha^{\prime-1}}(\alpha^\prime s)^n
    =\frac{s^2}{\Mpl^2}
    -g_1g_2\left(\alpha^{\prime}s\right)^3 t\,\Phi(\alpha^\prime s,1,1-\alpha^\prime t)\,,
\end{align}
where $\Phi$ is the Lerch zeta function.
By expanding around $t=0$, we obtain
\begin{align}
    -\left(\alpha^{\prime}s\right)^3 t\,\Phi(\alpha^\prime s,1,1-\alpha^\prime t)
    =-\alpha^\prime s^2\sum_{n=1}^\infty\left(\alpha^{\prime}t\right)^n\mathrm{Li}_n(\alpha^\prime s)
    \to \alpha^\prime s^2\sum_{n=1}^\infty\frac{\left(\alpha^{\prime}t\log(\alpha^\prime s)\right)^n}{n!}\,,
\end{align}
for $\alpha^\prime s\gg1$.
Here we have used
\begin{align}
    &\mathrm{Li}_n(\alpha^\prime s)=-\frac{\left(\log(\alpha^\prime s)\right)^n}{n!}
    +\mathcal{O}\left(\left(\log(\alpha^\prime s)\right)^{n-1}\right),
    &&\text{for $\alpha^\prime s\gg1$}\,.
\end{align}
Now, assuming that $g_1g_2\alpha^\prime=\Mpl^{-2}$, we obtain
\begin{align}
    \frac{s^2}{\Mpl^2}
    +g_1g_2\sum_{n=1}^\infty\frac{s^{2}t\,\alpha^{\prime}}{t-n\alpha^{\prime-1}}(s\alpha^\prime)^n
    &\to\frac{s^2}{\Mpl^2}
    +g_1g_2\alpha^\prime s^2\sum_{n=1}^\infty\frac{\left(\alpha^{\prime}t\log(\alpha^\prime s)\right)^n}{n!}
    \nonumber\\
    &=\frac{s^2}{\Mpl^2}e^{\alpha^{\prime}t\log(\alpha^\prime s)}=\frac{s^{2+\alpha^\prime t}}{\Mpl^2}\alpha^{\prime\alpha^\prime t}\,.
\end{align}
As a result, we obtain
\begin{align}
    &\frac{s^{2+\alpha^\prime t}}{\Mpl^2}\alpha^{\prime\alpha^\prime t}F(t,m^2)\,.
\label{eq:t-channel_sum}\end{align}
This is nothing but our proposal~\eqref{Eq:F_loop} for small $t$.
Although we are not able to compute the numerical factor of each diagram,\footnote{In general, $n$-dependent coefficient appears in the summation. Moreover, the couplings $g_{1,2}$ can depend on $n$. It is interesting to study the condition to realize the Reggeized amplitude.} this illustrates the idea of how Reggeization occurs, and how the parametrically large prefactor appears.


\section{Discussion}
\label{sec:discussion}
In this paper, we have studied the gravitational positivity bound at the one-loop level.
We first reviewed the gravitational positivity bound at the tree-level, where the $t$-channel pole corresponding to the graviton exchange can be canceled by assuming Regge behavior for the amplitude at high energy.
Next, we saw two examples of unitary, Reggeized gravitational amplitudes for which the potentially negative finite contribution to the positivity bounds can be found.
Finally, we moved to the features of Reggeized amplitudes at one loop. It is known that the one-loop EFT amplitude leads to the parametrically large negative contribution to the positivity bound. We argued for a form of the Reggeized amplitude at the one-loop level based on the analytic structure of the $t$-channel exchange diagram of the graviton and the higher spin particles in the Regge tower. 
We also provide an argument that the graviton $t$-channel pole can not be neglected by showing a counterexample to Eq.~\eqref{eq:lowCutOff} in string theory. These results are consistent with Refs.~\cite{Alberte:2021dnj,Herrero-Valea:2022lfd,deRham:2022gfe}.

As a future direction, we may consider a string theory setup to explicitly check the form of the amplitude at the one-loop of the matter fields.
Ideally, we want to analyze the amplitude in the Calabi-Yau compactification near the conifold point, but this is hard. Instead, we may consider the compactifications on the flat background.
For instance, we may take the non-supersymmetric $\SO(16)\times\SO(16)$ heterotic string and compactify it on $T^6$.\footnote{We can consider the same setup for the supersymmetric heterotic string theory.}
In 10d there are fermions whose representation is $({\bf16},{\bf16})+({\bf128},{\bf1})+({\bf1},{\bf128})$ under $\SO(16)\times\SO(16)$.
These will become Dirac fermions in 4d after compactification.
By turning on a VEV for the Wilson lines, the gauge symmetry is broken from $\SO(16)\times\SO(16)$ to $\U(1)^{16}$.\footnote{In addition, there are KK and winding $\U(1)$s, but these are not important here.}
Then the $({\bf16},{\bf16})$ fermions have charge
\begin{equation}
    \big(\underbrace{1,0,\ldots,0}_{\U(1)^8},\underbrace{1,0,\ldots,0}_{\U(1)^8}\big)
\end{equation}    
(and its permutations) under $\U(1)^{16}$.  Similarly, $({\bf128},{\bf1})$ has charge
\begin{equation}
    \big(\underbrace{\pm\tfrac{1}{2},\ldots,\pm\tfrac{1}{2}}_{\U(1)^8},\underbrace{0,\ldots,0}_{\U(1)^8}\big) \,.
\end{equation}
These fermions receive masses from the Wilson line, and the masses are chosen to be arbitrary values. 
This setup is close to the non-supersymmetric QED coupled with light matters, and loops of these fermions may give a negative contribution to the coefficient of $s^2$. On the other hand, the typical mass of the higher-spin particles is always the string scale, independent of the VEV of the Wilson line.
It is an interesting task to compute the one-loop string amplitude at $\mathcal{O}(\Mpl^{-2})$ in this setup to check the details of our proposal in Sec.~\eqref{sec:reggeloop}.

\newpage
\normalsize

\subsection*{Acknowledgments}
The work of Y.H.\ and G.L.\ is supported in part by MEXT Leading Initiative for Excellent Young Researchers Grant Number JPMXS0320210099.
The work of S.N.\ is supported in part by JSPS KAKENHI Grant Number 21J15497.

\appendix

\section{Notation}
\label{sec:notation}

We follow the notation in Ref.~\cite{Alberte:2020bdz}.
We consider the scattering $\gamma_1\gamma_2\to\gamma_3\gamma_4$.
The momenta $k^\mu_{1,2,3,4}$ are parametrized as (all-ingoing notation)
\begin{align}
    &k^\mu_1=(k,0,0,k)\,,
    &&k^\mu_2=(k,0,0,-k)\,,
    \nonumber\\
    &k^\mu_3=-(k,k\sin\theta,0,k\cos\theta)\,,
    &&k^\mu_4=-(k,-k\sin\theta,0,-k\cos\theta)
\end{align}
We define $\epsilon^\mu_{1,2,3,4}$ as a polarization vector of the external photon. The $\pm$ polarization corresponds to
\begin{align}
    &\epsilon^\mu_1(\pm)=\frac{1}{\sqrt{2}}(0,1,\pm1,0)\,,
    &&\epsilon^\mu_2(\pm)=\frac{1}{\sqrt{2}}(0,-1,\pm1,0)\,,
    \nonumber\\
    &\epsilon^\mu_3(\pm)=\frac{1}{\sqrt{2}}(0,\cos\theta,\pm i,-\sin\theta)\,,
    &&\epsilon^\mu_4(\pm)=\frac{1}{\sqrt{2}}(0,\cos\theta,\pm i,\pm\sin\theta)\,.
\end{align}
In all-ingoing notation, the amplitude is written as
\begin{align}
    \A(h_1,h_2,h_3,h_4;k_1,k_2,k_3,k_4)=\epsilon_1^\mu(h_1)\epsilon_2^\nu(h_2)\epsilon_3^\alpha(h_3)\epsilon_4^\beta(h_4)\A^{\mu\nu\alpha\beta}(k_1,k_2,k_3,k_4)\,,
\end{align}
where $h_{1,2,3,4}=\pm1$ is the helicity of the external photons in all-incoming notation.
$\epsilon_{ij}$ is defined as
\begin{align}
    \epsilon_{ij}:=\epsilon_i\cdot\epsilon_j
\end{align}
Explicitly, we obtain
\begin{align}
    &\epsilon_{12}=-\frac{1}{2}-\frac{h_1h_2}{2}\,,
    &&\epsilon_{13}=-\frac{h_1h_3}{2}+\frac{1}{2}+\frac{t}{s}\,,
    &&\epsilon_{14}=-\frac{h_1h_4}{2}-\frac{1}{2}-\frac{t}{s}\,,
    \nonumber\\
    &\epsilon_{34}=-\frac{1}{2}-\frac{h_3h_4}{2}\,,
    &&\epsilon_{24}=-\frac{h_2h_4}{2}+\frac{1}{2}+\frac{t}{s}\,,
    &&\epsilon_{23}=-\frac{h_2h_3}{2}-\frac{1}{2}-\frac{t}{s}\,,
\end{align}
The inner products between external momenta and polarizations are
\begin{align}
    &k_1\cdot \epsilon_3=k_3\cdot \epsilon_1=k_2\cdot \epsilon_4=k_4\cdot \epsilon_2=-\frac{\sqrt{tu}}{\sqrt{2s}}\,,\nonumber\\
    &k_1\cdot \epsilon_4=k_4\cdot \epsilon_1=k_2\cdot \epsilon_3=k_3\cdot \epsilon_2=\frac{\sqrt{tu}}{\sqrt{2s}}\,,\nonumber\\
    &\mathrm{(others)}=0\,.
\label{Eq:k_epsilon}\end{align}
The inner products among external momenta are
\begin{align}
    &k_1\cdot k_2=k_3\cdot k_4=s/2\,,
    &&k_1\cdot k_3=k_2\cdot k_4=t/2\,,
    \nonumber\\
    &k_1\cdot k_4=k_2\cdot k_3=u/2\,,
    &&k_1^2=k_2^2=k_3^2=0\,.
\end{align}
where $(s,t,u)$ are the Mandelstam variables:
\begin{align}
&s=(k_1+k_2)^2=2k_1\cdot k_2,
&&t=(k_1+k_3)^2=2k_1\cdot k_3,
&&u=(k_1+k_4)^2=2k_1\cdot k_4\,.
\end{align}

\section{Unimportant diagrams}
\label{sec:diagrams}

\begin{figure}[t]
    \begin{center}
        \includegraphics[width=75mm, height=56.25mm]{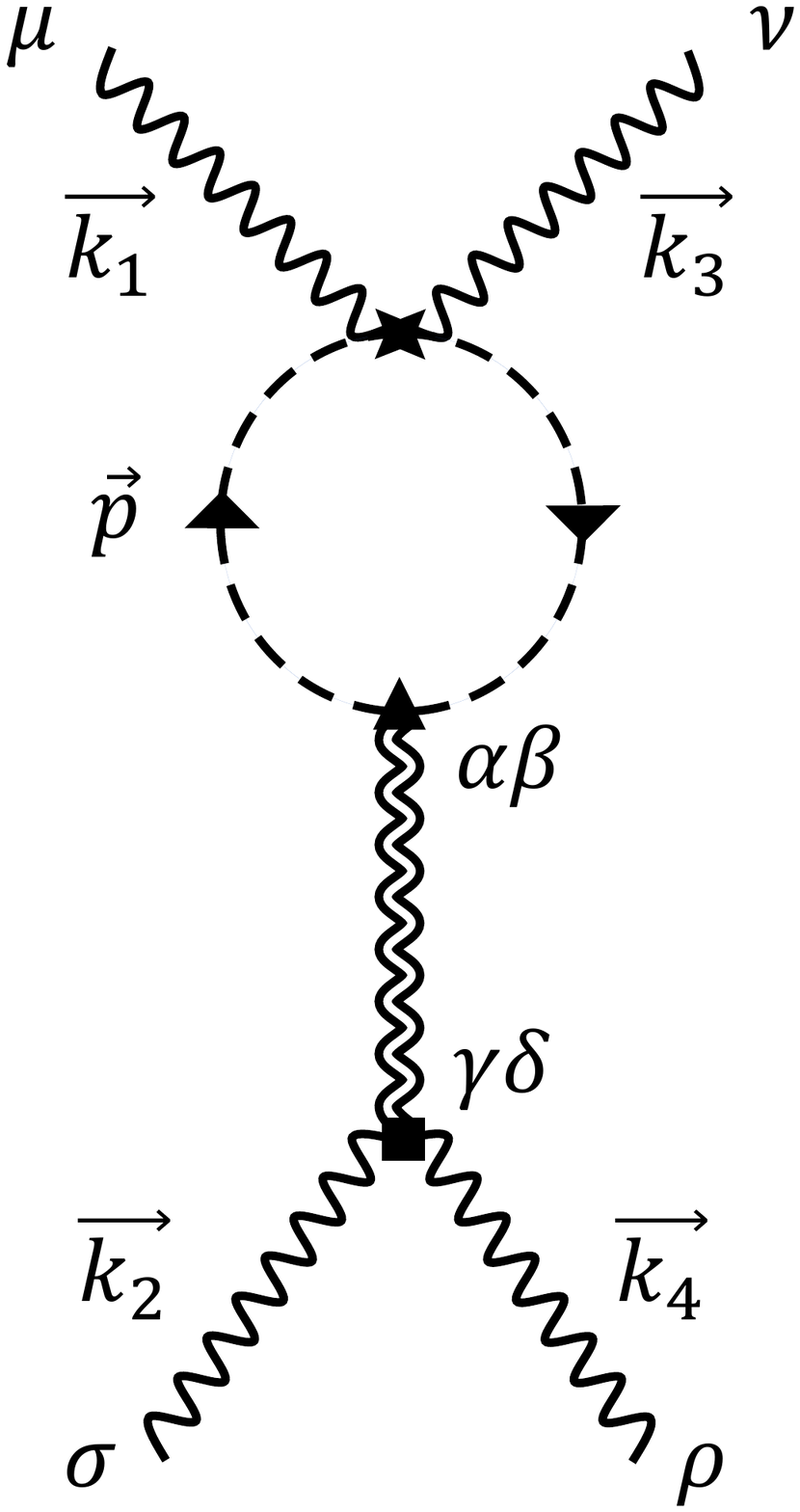}
        \includegraphics[width=75mm, height=56.25mm]{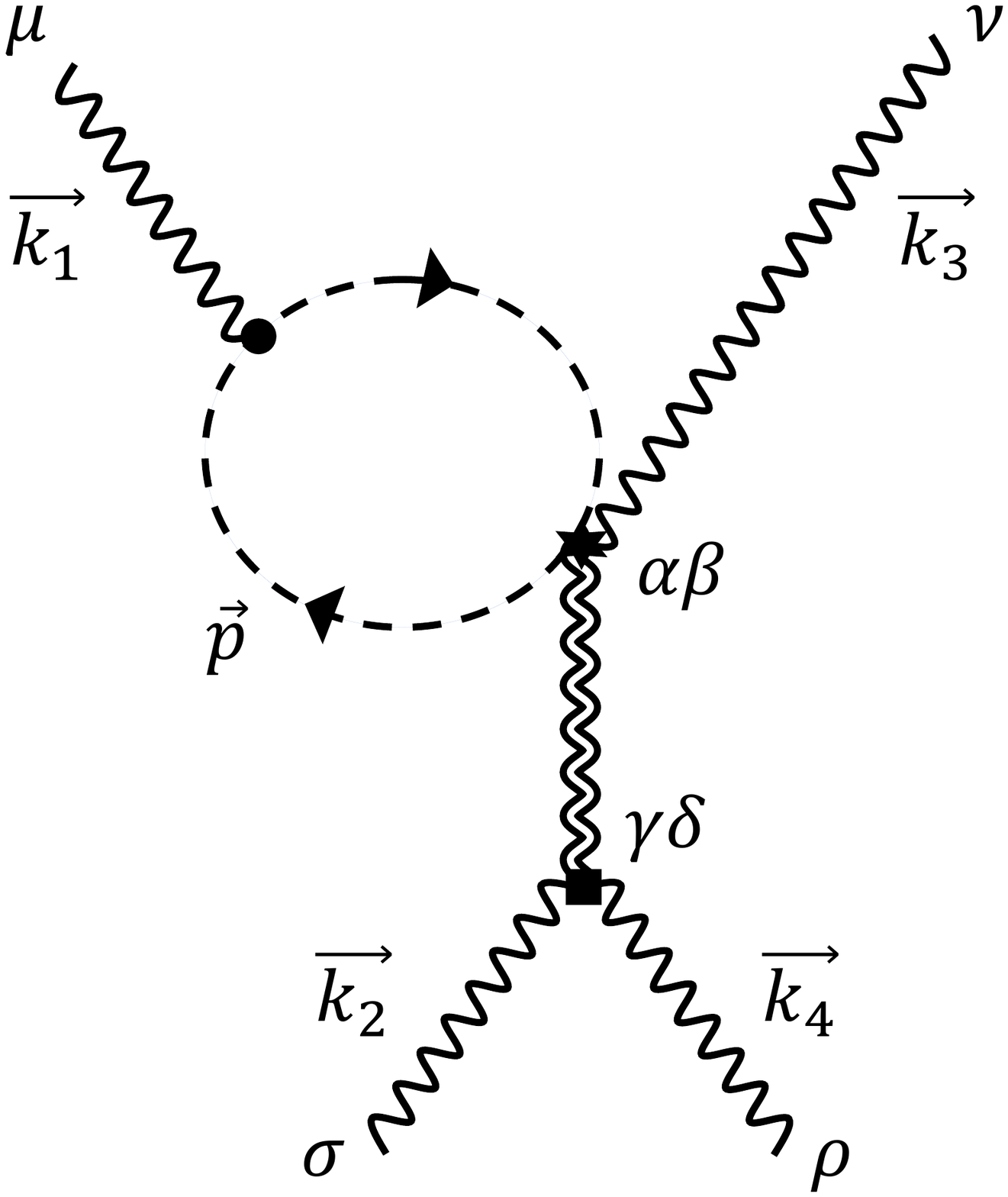}
        \includegraphics[width=75mm, height=56.25mm]{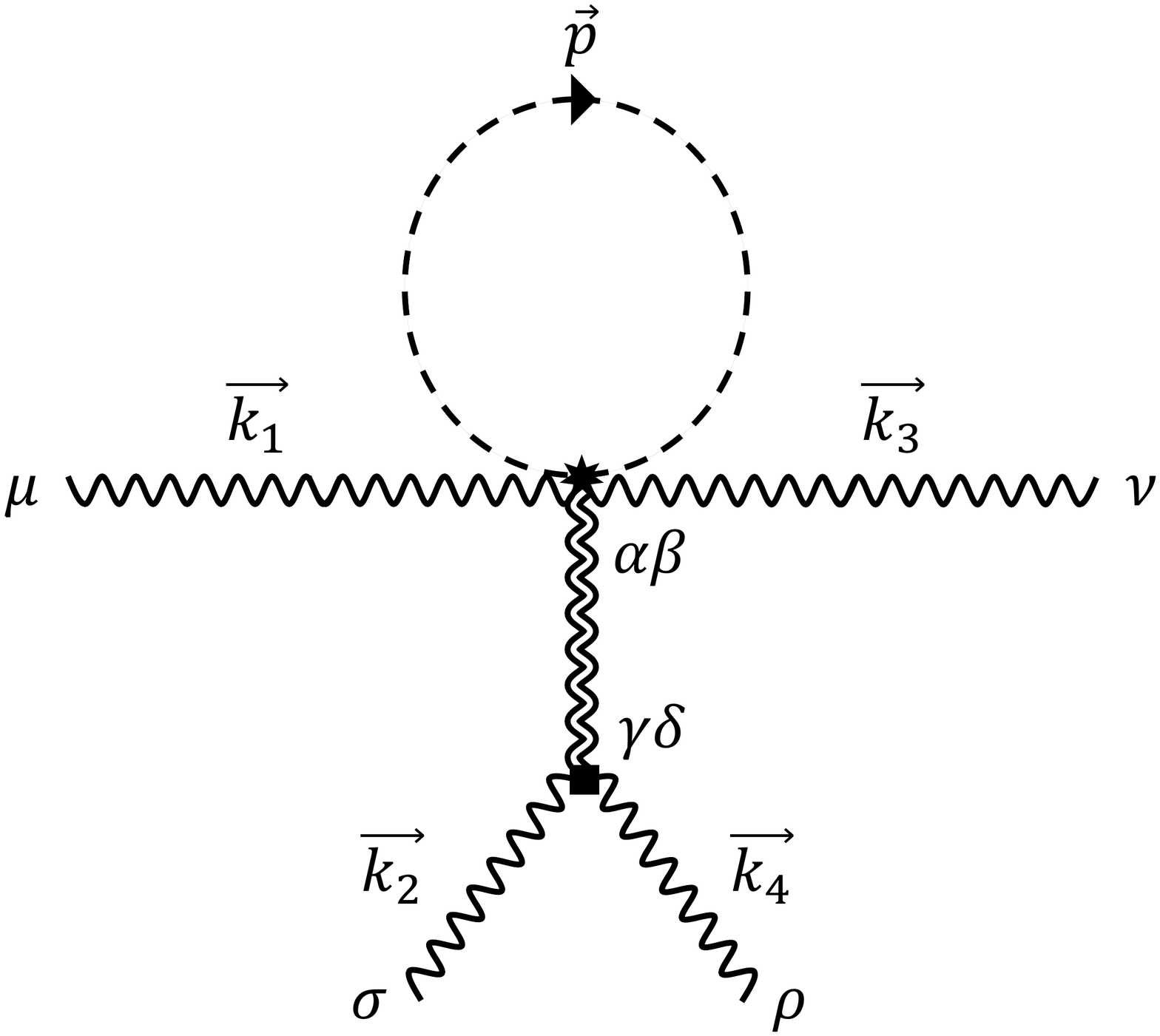}
        \caption{The diagrams whose contribution does not contain the light mass parameter $m^2$ in the denominator for $t\to0$.}
        \label{fig:diagrams}
    \end{center}
\end{figure}

In the main text, we have considered the diagrams in Fig.~\ref{fig:diagram}.
However, there are other diagrams in Fig.~\ref{fig:diagrams}, which are potentially relevant.
These diagrams are not relevant to us because these contributions do not contain the light mass parameter $m^2$ in the denominator for $t\to0$.
This can be seen from the power counting of the loop integral.
All the diagrams in Figs.~\ref{fig:diagrams} contain two scalar propagators, which scale as $p^{-4}$, where $p$ is the loop integral.
Therefore, after integrating over $p$, we get $\log(m^2)$ at most.
The derivative interaction at the vertex does not change the conclusion because it changes the scaling of the integrand as $p^{-4+n}$ where $n\geq0$.

\section{Reggeization at the tree-level}\label{sec:tree_Regge}
In this appendix, we show how the tree-level amplitude is Reggeized starting from the pole expansion~\eqref{eq:pole_expansion}, along the line with the end of Sec.~\ref{sec:oneloop}.\footnote{Of course, we know the full expression~\eqref{Eq: Virasoto-Shapiro}. It is easy to show the Regge behavior from Eq.~\eqref{Eq: Virasoto-Shapiro}. Nevertheless, it is instructive to see how the same result emerges from the pole expansion.}
In the following, we assume the limit $s\to \pm i\infty$.\footnote{The argument is not valid for other directions in $s$-plane.}
For large $s$, Eq.~\eqref{eq:pole_expansion} is written as
\begin{align}
    \A(s,t) &\sim -\frac{4Ps^4}{\alpha^\prime}\sum_{n=0}^{\infty} 
    \frac{1}{(n!)^2}\left(\frac{\alpha^\prime s}{4}\right)^{2n-2}
    \left(
    \frac{1}{t-4n\alpha^{\prime-1} } 
    - 
    \frac{1}{t+s+4n\alpha^{\prime-1}}
    \right)
    \nonumber\\
    &=-\frac{64P s^2}{\alpha^{\prime3}}\left[\left(
    \frac{1}{t} - \frac{1}{t+s}
    \right)
    +\sum_{n=1}^{\infty} 
    \frac{1}{(n!)^2}\left(\frac{\alpha^\prime s}{4}\right)^{2n}
    \left(
    \frac{1}{t-4n\alpha^{\prime-1} } 
    - 
    \frac{1}{t+s+4n\alpha^{\prime-1}}
    \right)\right]\,.
\label{eq:pole_larges}\end{align}
We expand the second term around $t=0$, and then perform the summation from $n=1$ to $n=\infty$.
By keeping the leading term for $s\to\infty$ in each order of $t$, we obtain 
\begin{align}
    \sum_{n=1}^{\infty} 
    \frac{1}{(n!)^2}\left(\frac{\alpha^\prime s}{4}\right)^{2n}
    \left(
    \frac{1}{t-4n\alpha^{\prime-1}} 
    - 
    \frac{1}{t+s+4n\alpha^{\prime-1}}
    \right)
    \sim
    \frac{\alpha^\prime}{4}\sum_{m=1}^{\infty} \frac{1}{m!}\left(\frac{\alpha^\prime t}{4}\right)^{m-1}\left(\log\left(\frac{\alpha^{\prime2}s^2}{16}\right)\right)^m\,.
\end{align}
By substituting this into Eq.~\eqref{eq:pole_larges}, we obtain the Reggeized amplitude:
\begin{align}
    \A(s,t)&\sim -\frac{16Ps^2}{\alpha^{\prime2}}\sum_{m=0}^{\infty} \frac{1}{m!}\left(\frac{\alpha^\prime t}{4}\right)^{m-1}\left(\log\left(\frac{\alpha^{\prime2}s^2}{16}\right)\right)^m
    +\frac{64P s^2}{\alpha^{\prime3}(t+s)}
    \nonumber\\
    &=-\frac{64Ps^2}{\alpha^{\prime3}}\left(\frac{\alpha^{\prime2}s^2}{16}\right)^{\alpha^{\prime}t/4}
    +\frac{64P s^2}{\alpha^{\prime3}(t+s)}\,.
\end{align}

\bibliographystyle{JHEP}
\bibliography{Bibliography}

\end{document}